\UseRawInputEncoding
\documentclass[preprint2]{aastex631}

\received{August 28, 2023}
\revised{April 2, 2024}
\accepted{April 19, 2024}

\shorttitle{SN 2023ixf Progenitor Properties}
\shortauthors{Van Dyk et al.}

\begin{document}

\title{The SN 2023\lowercase{ixf} Progenitor in M101: II. Properties}

\correspondingauthor{Schuyler D.~Van Dyk}
\email{vandyk@ipac.caltech.edu}

\author[0000-0001-9038-9950]{Schuyler D.~Van Dyk}
\affiliation{Caltech/IPAC, Mailcode 100-22, Pasadena, CA 91125, USA}

\author[0000-0002-2996-305X]{Sundar~Srinivasan}
\affiliation{Instituto de Radioastronom{\'i}a y Astrof{\'i}sica, UNAM, Antigua Carretera a P\'atzcuaro 8701, Ex-Hda. San Jos\'e de la Huerta,  Morelia 58089, Mich., Mexico}

\author[0000-0003-0123-0062]{Jennifer E.~Andrews}
\affiliation{Gemini Observatory/NSFÕs NOIRLab, 670 N. AÕohoku Place, Hilo, HI 96720, USA}

\author[0000-0001-6360-992X]{Monika Soraisam}
\affiliation{Gemini Observatory/NSFÕs NOIRLab, 670 N. AÕohoku Place, Hilo, HI 96720, USA}

\author[0000-0003-4610-1117]{Tam\'as Szalai}
\affiliation{Department of Experimental Physics, Institute of Physics, University of Szeged, D{\'o}m t{\'e}r 9, 6720 Szeged, Hungary}
\affiliation{ELKH-SZTE Stellar Astrophysics Research Group, Szegedi {\'u}t, Kt. 766, 6500 Baja, Hungary}
\affiliation{MTA-ELTE Lend\"ulet "Momentum" Milky Way Research Group, Hungary}

\author[0000-0002-2532-2853]{Steve B.~Howell}
\affiliation{NASA Ames Research Center, Moffett Field, CA 94035, USA}

\author[0000-0002-0531-1073]{Howard Isaacson}
\affiliation{Department of Astronomy, University of California, Berkeley, CA 94720-3411, USA}

\author[0000-0001-6685-0479]{Thomas Matheson}
\affiliation{NSF's National Optical-Infrared Astronomy Research Laboratory, 950 North Cherry Avenue, Tucson, AZ 85719, USA}

\author[0000-0003-0967-2893]{Erik Petigura}
\affiliation{Department of Physics \& Astronomy, University of California Los Angeles, Los Angeles, CA 90095, USA}

\author[0000-0002-1161-3756]{Peter~Scicluna}
\affiliation{European Southern Observatory, Alonso de Cordova 3107, Santiago RM, Chile}
\affiliation{Space Science Institute, 4750 Walnut Street, Suite 205, Boulder, CO 80301, USA}

\author[0000-0002-4434-2307]{Andrew W.~Stephens}
\affiliation{Gemini Observatory/NSFÕs NOIRLab, 670 N. AÕohoku Place, Hilo, HI 96720, USA}

\author[0000-0002-4290-6826]{Judah Van Zandt}
\affiliation{Department of Physics \& Astronomy, University of California Los Angeles, Los Angeles, CA 90095, USA}

\author[0000-0002-2636-6508]{WeiKang Zheng}
\affiliation{Department of Astronomy, University of California, Berkeley, CA 94720-3411, USA}

\author[0000-0002-6154-7558]{Sang-Hyun~Chun}
\affiliation{Korea Astronomy and Space Science Institute, 776 Daedeokdae-ro, Yuseong-gu, Daejeon 34055, Republic of Korea}

\author[0000-0003-3460-0103]{Alexei V.~Filippenko}
\affiliation{Department of Astronomy, University of California, Berkeley, CA 94720-3411, USA}

\begin{abstract}
We follow our first paper with an analysis of the ensemble of the extensive pre-explosion ground- and space-based infrared observations of the red supergiant (RSG) progenitor candidate for the nearby core-collapse supernova SN 2023ixf in Messier 101, together with optical data prior to explosion obtained with the {\sl Hubble Space Telescope\/} ({\sl HST}). We have confirmed the association of the progenitor candidate with the SN, as well as constrainted the metallicity at the SN site, based on SN observations with instruments at Gemini-North. The internal host extinction to the SN has also been confirmed from a high-resolution Keck spectrum. We fit the observed spectral energy distribution (SED) for the star, accounting for its intrinsic variability, with dust radiative-transfer modeling, which assume a silicate-rich dust shell ahead of the underlying stellar photosphere. The star is heavily dust-obscured, likely the dustiest progenitor candidate yet encountered. We found median estimates of the star's effective temperature and luminosity of 2770~K and $9.0 \times 10^4\ L_{\odot}$, with 68\% credible intervals of 2340--3150~K and (7.5--10.9) $\times 10^4\ L_{\odot}$. The candidate may have a Galactic RSG analog, IRC~$-10414$, with a strikingly similar SED and luminosity. Via comparison with single-star evolutionary models we have constrained the initial mass of the progenitor candidate from $12\ M_{\odot}$ to as high as $14\ M_{\odot}$. We have had available to us an extraordinary view of the SN 2023ixf progenitor candidate, which should be further followed up in future years with {\sl HST\/} and the {\sl James Webb Space Telescope}.
\end{abstract}

%% Keywords should appear after the \end{abstract} command. 
%% The AAS Journals now uses Unified Astronomy Thesaurus concepts:
%% https://astrothesaurus.org
%% You will be asked to selected these concepts during the submission process
%% but this old "keyword" functionality is maintained in case authors want
%% to include these concepts in their preprints.


\bibpunct[;]{(}{)}{;}{a}{}{;}

\section{Introduction} \label{sec:intro}

Among the supernovae (SNe) that arise from the collapse of the cores of massive (initial mass $M_{\rm ini} \gtrsim 8\ M_{\odot}$) stars, nearly half locally are of Type II \citep{Smith2011} --- those which show hydrogen lines in their near-maximum-light optical spectra. The overwhelming majority of SNe~II further exhibit extended plateaus in their light curves and are referred to as SNe~II-P. We now have compelling observational evidence, supporting the theoretical expectations, that SNe II-P are the explosions of massive stars in the final red supergiant (RSG) phase of their evolution; in particular, the direct identifications have been made in more than 20 cases of RSGs as the progenitors of SNe~II-P (e.g., \citealt{Smartt2009,Smartt2015,VanDyk2017}). Most of these direct detections so far have been of relatively low $M_{\rm ini}$ ($\sim 8$--$10\ M_{\odot}$), consistent with the low-luminosity events resulting from their explosions \citep[e.g.,][]{Smartt2004,Maund2005,VanDyk2023b}. \citet{Smartt2009} were the first to point out that a ceiling might exist on the maximum possible $M_{\rm ini}$ for SN~II-P progenitors, at $\sim 17\ M_{\odot}$ (although \citealt{Davies2020} have since argued that this limit could extend as high as $\approx 20\ M_{\odot}$). That some Galactic RSGs appear to have $M_{\rm ini} \lesssim 25$--$30\ M_{\odot}$ is perplexing, although the presumed progenitor mass ceiling is still to be adequately challenged observationally (\citealt{Davies2020} argue that the progenitor sample needs to be at least doubled). Every new example of a possible progenitor identification is therefore welcome and valuable.

The known RSG progenitors of SNe II-P have been characterized so far, mostly (although not exclusively) based on detections in more than one photometric band, such that a spectral energy distribution (SED) can be constructed and analyzed. This has been accomplished primarily via serendipitous pre-explosion imaging data obtained with the {\sl Hubble Space Telescope\/} ({\sl HST}), enhanced in a few cases with further detections in data from the {\sl Spitzer Space Telescope}. Solely ground-based detections have been far less common, e.g., SN~2008bk \citep{Mattila2008,VanDyk2012a}, SN~2012A \citep{Tomasella2013}, and the most famous example, of course, SN 1987A (although the progenitor was a {\it blue\/} supergiant, not a red one; \citealt{West1987,Sonneborn1987,Gilmozzi1987}). In the lower-luminosity, lower-mass cases the observed SED can be fit reasonably well with a bare photospheric model, such as SN~2008bk \citep{ONeill2021} and SN~2018aoq \citep{ONeill2019}. However, other cases in which the progenitor was inevitably inferred to be of higher luminosity, higher $M_{\rm ini}$, also required the presence of circumstellar dust to be accounted for in modeling the SED, as for SN~2012aw \citep{VanDyk2012b,Kochanek2012,Fraser2012} and SN~2017eaw \citep{Kilpatrick2018,VanDyk2019,Rui2019}.

That this is true is not surprising, given our knowledge of RSGs locally. For Local Group RSGs, the more luminous the RSG, the higher is its inferred mass-loss rate ($\dot M$), and the dustier is its circumstellar environment \citep{Massey2005,Verhoelst2009,Bonanos2010,Mauron2011,Yang2018}. As we pointed out in \citet[][ Paper I hereafter]{Soraisam2023b}, RSGs are also known to show pulsationally-driven, semiregular variability \citep[e.g.,][]{Heger1997,Kiss2006,Soraisam2018,Chatys2019}. Additionally, for instance, in a study of Large Magellanic Cloud RSGs \citet{Yang2018} found not only correlations between luminosity and $\dot M$, and luminosity and variability, but also between variability and $\dot M$; in short, the more luminous RSG population exhibited a much larger infrared (IR) excess, $\dot M$, and variability than did the less luminous objects.

Radial pulsations of RSGs can affect the structure of the star's envelope, which further affects the luminosity evolution of the SN explosion, in that more luminous events tend to decline more steeply on the plateau, and any nonradial pulsations could also lead to observed asymmetries during the plateau phase \citep{Goldberg2020}. Additionally, injection of even a small amount ($\sim 10^{46}$--$10^{47}$ erg) of deposited energy into the envelope, possibly in the form of internal waves excited by late-stage core convection \citep[e.g.,][]{Shiode2014,Fuller2017}, can drive ejection of dense and confined circumstellar matter (CSM) prior to explosion, leading to hot, luminous early-time emission and short-lived ``flash-ionization'' spectral features \citep{Morozova2020}. This energy injection might manifest itself as pre-SN outbursts shortly before explosion, which was problematic for, e.g., SN 2017eaw, for which no such luminous outbursts were detected in the IR \citep{VanDyk2019,Tinyanont2019}. However, such outbursts may be more energetic and pronounced in lower-mass ($M_{\rm ini} \lesssim 12\ M_{\odot}$) progenitors \citep{Wu2021} and may affect the optical luminosity more than the IR \citep{Davies2022}. Furthermore, \citet{Beasor2021} concluded that, if some mechanism  at late nuclear burning phases led to instabilities and episodic mass loss, then a lower, steady $\dot M$ at earlier phases \citep{Beasor2020} would potentially leave the RSG with more envelope to lose in the final years of its life, leading to more CSM at core collapse.

This, then, brings us to SN~2023ixf in the famous nearby ``Pinwheel Galaxy'' Messier 101 (M101; NGC 5457). Given its proximity and brightness, SN~2023ixf has already achieved a level of fame of its own and has been considered the ``SN of the decade,'' at least in the early years of the 2020s. Much has already been written about this SN, and we briefly summarize here only a portion of that. Within hours of its discovery by \citet{Itagaki2023} on 2023 May 19.727 (UTC is adopted throughout this paper), \citet{Perley2023} classified the SN as Type~II, showing a strong blue continuum and prominent optical flash-ionization features of H, He~{\sc i}/{\sc ii}, N~{\sc iii}/{\sc iv}, and C~{\sc iii}/{\sc iv}, all indicative of the presence of CSM. Early photometric and spectroscopic studies of the SN have described the evidence for interaction of the SN shock with a dense, confined ($<2 \times 10^{15}$~cm) CSM, which further boosted (by $\gtrsim 2$ mag) the SN's early-time optical luminosity \citep{Yamanaka2023,Jacobson-Galan2023,Hosseinzadeh2023,Smith2023,Bostroem2023,Teja2023,Hiramatsu2023}. \citet{Zimmerman2024} further constrained the confined CSM, which extended shock breakout, at $<2 \times 10^{14}$~cm. \citet{Smith2023}, \citet{Vasylyev2023}, and \citet{Li2023} presented evidence for asymmetry in the CSM and the SN ejecta. Further signs of early CSM interaction at other wavelengths include the detection of the SN in X-rays (up to 20~keV) at several epochs, starting about four days after explosion and beyond \citep{Grefenstette2023,Mereminskiy2023,Chandra2023b,Chandra2024}, as well as, after initial radio and submillimeter nondetections \citep{Chandra2023a,Berger2023}, detection at cm wavelengths $\sim 29$~d post-discovery \citep{Matthews2023}. Early attempts were made to detect the SN in $\gamma$-rays \citep{Ravensburg2024} and neutrinos \citep{Thwaites2023,Guetta2023}, with null results. More recent, extended photometric and spectroscopic monitoring (\citealt{Bianciardi2023}; W.~Zheng et al., in prep.) appears to indicate that SN~2023ixf is a short-plateau SN~II-P or an SN~II-P/II-Linear hybrid.

A number of constraints on the explosion epoch have been presented by both amateur and professional astronomers, which \citet{Yaron2023} initially summarized. \citet{Hosseinzadeh2023} also performed a careful analysis of the various constraints and narrowed the date of explosion to MJD 60082.75 (2023 May 18.75), which we adopted in Paper~I and do so here as well.

SN 2023ixf occurred just on the outskirts of the giant H~{\sc ii} region NGC 5461 (one of the five largest and brightest in M101; \citealt{Seyfert1940}), only $24{\farcs}8$ west and $23{\farcs}0$ south of the region's center. It was therefore relatively straightforward to identify a progenitor candidate for SN~2023ixf, especially at {\sl Spitzer\/} IR wavelengths (\citealt{Szalai2023,Kilpatricketal2023,Jencson2023,Niuetal2023}; Paper~I), based solely on the SN's reported absolute position \citep{Itagaki2023}. As \citet{Kilpatricketal2023}, \citet{Jencson2023}, and Paper~I found, the candidate was also detectable in various pre-SN near-IR imaging data. A faint possible counterpart was detectable in the {\sl HST\/} F814W ($\sim I$) band as well \citep{Soraisam2023a,Pledger2023,Kilpatricketal2023,Jencson2023,Niuetal2023}. All of the available photometric information pointed to an RSG for the candidate. \citet{Kilpatricketal2023}, \citet{Jencson2023}, \citet{Niuetal2023}, and Paper~I all detailed the variability of the candidate in the IR. In Paper~I we found a fundamental period of the variability of $\sim 1091$~d, implying a long-period, semiregular nature. Based on the {\sl HST\/} data alone, \citet{Pledger2023} estimated a low $M_{\rm ini} \approx 8$--$10\ M_{\odot}$, whereas from modeling of the combined {\sl HST\/} and {\sl Spitzer\/} data \citet{Kilpatricketal2023} found a higher bolometric luminosity, $L_{\rm bol}=10^{4.74}\ L_{\odot}$, and $M_{\rm ini} = 11\ M_{\odot}$. \citet{Jencson2023} constrained the luminosity and initial mass at even higher values, $10^{5.1}\ L_{\odot}$ and $M_{\rm ini} = 17 \pm 4\ M_{\odot}$. \citet{Niuetal2023} also found similar values. In Paper~I, from the RSG period-luminosity relation we provided an estimate, also on the high side, of $M_{\rm ini} = 20 \pm 4\ M_{\odot}$.

As a companion paper to our analysis of the progenitor candidate variability, we present here our assessment of the overall properties of the candidate. We summarize in Section~\ref{sec:observations}  all of the available data that we have collected and analyzed. In Section~\ref{sec:progID} we make a precise association between the young SN and the candidate in the {\sl HST\/} and near-IR data. We compile the available host-galaxy distances and adopt a value (and uncertainty) in Section~\ref{sec:distance}.  We provide in Section~\ref{sec:extinction} our estimate for the total reddening to the SN, and in Section~\ref{sec:metallicity} we place constraints on the metallicity at the SN site, based on a nearby H~{\sc ii} region. In Section~\ref{sec:sedfit} we describe our model fitting to the candidate's SED, while in Section~\ref{sec:properties} we analyze that fit to provide an estimate of the candidate's properties.  We summarize and discuss our results in Section~\ref{sec:discussion}.

We note that M101 is nearly face-on (inclination $8\arcdeg$; \citealt{Jarrett2003}). We assume throughout a redshift for the galaxy of $z=0.000804$ ($+241.0$ km s$^{-1}$; via the NASA/IPAC Extragalactic Database). M101 has been the host of SN~1909A and SN~1951H (both unclassified), as well as SN~II~1970G (at the edge of another giant H~{\sc ii} region, NGC 5455; e.g., \citealt{Winzer1974,Fesen1993}) and the well-studied Type Ia SN~2011fe (e.g., \citealt{Parrent2012,Tucker2022}). 

\section{Observations}\label{sec:observations}

Given the proximity of M101 and its nearly face-on orientation, the galaxy and, in particular, NGC 5461 have been targets of study for decades by a slew of ground- and space-based facilities. In turn, for similar reasons SN 2023ixf has already been intently observed by a growing number of investigators. Here we summarize the observational data that we considered for this study.

\subsection{{\sl Hubble Space Telescope\/} Imaging}

The SN 2023ixf site was imaged serendipitously in a number of bands by {\sl HST\/} over nearly 24~yr prior to explosion; see Table~\ref{tab:hst_obs} for a listing. The specific observations analyzed can be accessed via the Mikulski Archive for Space Telescopes (MAST): \dataset[10.17909/j2aw-rp24]{http://dx.doi.org/10.17909/j2ax-4p24}. The fields were observed with the Wide Field and Planetary Camera 2 (WFPC2), the Advanced Camera for Surveys Wide-Field Channel (ACS/WFC), and the Wide Field Camera 3 UVIS channel (WFC3/UVIS).

We processed each band in each dataset individually, first by mosaicking the individual frames using {\tt Astrodrizzle} \citep{STScI2012}. An additional benefit to making the mosaics in this way is that cosmic-ray (CR) hits are flagged in the Data Quality layer of each frame. We then ran the individual frames through {\tt Dolphot} \citep{Dolphin2016} to extract photometric measurements via point-spread-function (PSF) fitting. For all of the data we adopted {\tt Dolphot} parameter values {\it FitSky}=3 and {\it RAper}=8, {\it InterpPSFlib}=1, with charge-transfer efficiency (CTE) correction set to false for the ACS and WFC3 data (set to true for the WFPC2 data). The CR-hit flagging is important, allowing for accurate aperture corrections to the PSF fitting to be estimated and applied. 

\begin{deluxetable*}{cccccc}
\tablewidth{0pt}
\tablecolumns{6}
\tablecaption{{\sl HST\/} Observations Containing the Progenitor Candidate\label{tab:hst_obs}}
\tablehead{\colhead{Band} & \colhead{UT Date} & \colhead{Instrum.} & \colhead{Exp.~time} & \colhead{Mag} & \colhead{ProgID} \\
\colhead{} & \colhead{} & \colhead{} & \colhead{(s)} & \colhead{(Vega)} & \colhead{}}
\startdata
    F547M & 1999 Mar 24 & WFPC2     & 1400 & $> 25.7$    &  6829 \\
    F656N & 1999 Mar 23 & WFPC2     & 1360 & $> 21.7$    &  6829 \\
    F675W & 1999 Mar 24 & WFPC2     &  800 & 24.47(0.20) &  6829 \\
    F547M & 1999 Jun 17 & WFPC2     & 1000 & $> 25.7$    &  6829 \\
    F656N & 1999 Jun 16 & WFPC2     & 1200 & $> 21.7$    &  6829 \\
    F435W & 2002 Nov 16 & ACS/WFC   &  900 & $> 27.0$    &  9490 \\
    F555W & 2002 Nov 16 & ACS/WFC   &  720 & $> 26.7$    &  9490 \\
    F814W & 2002 Nov 16 & ACS/WFC   &  720 & 24.31(0.05) &  9490 \\
    F336W & 2003 Aug 27 & WFPC2     & 2400 & $> 23.5$    &  9720 \\
    F658N & 2004 Feb 10 & ACS/WFC   & 2440 & 24.53(0.17) &  9720 \\
    F502N & 2014 Mar 19 & WFC3/UVIS & 1310 & $> 24.5$     & 13361 \\
    F673N & 2014 Mar 19 & WFC3/UVIS & 1310 & 24.91(0.25) & 13361 \\
    F435W & 2018 Mar 30 & ACS/WFC   & 7423 & $> 28.6$    & 15192 \\
    F658N & 2018 Mar 30 & ACS/WFC   & 5770 & $> 25.4$    & 15192 \\ 
\enddata
\tablecomments{Columns: {\sl HST\/} observation band, mean UTC observation date, {\sl HST\/} instrument, total exposure time, {\tt Dolphot} magnitude or inferred 5$\sigma$ upper limit; {\sl HST\/} proposal ID number. Uncertainties in the {\tt Dolphot} magnitudes are in parentheses. Upper limits are estimates from formal $5\sigma$ detections via {\tt Dolphot}.}
\end{deluxetable*}

\subsection{{\sl Spitzer Space Telescope\/} Imaging}

In addition to the data obtained by {\sl Spitzer\/} with the IR Array Camera \citep[IRAC;][]{Fazio2004} at 3.6 and 4.5~$\mu$m that we described in Paper I, during the {\sl Spitzer\/} cryogenic mission the SN field was also observed on 2004 March 8 (program ID [PID] 60) at 5.8 and 8.0~$\mu$m. The field was also observed with the Multi-Band Imaging Photometer for Spitzer \citep[MIPS;][]{Rieke2004} in Scan mode on 2004 May 10 and 11 (PID 60), 2008 January 4 and 5 (PID 40352), and 2008 April 16 and 17 (PID 40352) at 24, 70, and 160~$\mu$m, as well as in Phot mode on 2006 May 7 (PID 20321) at 24~$\mu$m only.

\subsection{{\sl Herschel Space Observatory\/} Imaging}

The SN site was also serendipitously captured by the {\sl Herschel Space Observatory\/} Photodetector Array Camera and Spectrometer (PACS) at 70 and 160 $\mu$m on 2009 August 30 (program Calibration{\textunderscore}pvpacs{\textunderscore}24) and 2010 June 16 and 17 (KPOT{\textunderscore}rkennicu{\textunderscore}1), as well as with the Spectral and Photometric Imaging Receiver (SPIRE) at 250, 350, and 500 $\mu$m on 2009 December 30 (program KPOT{\textunderscore}rkennicu{\textunderscore}1).

\subsection{{\sl AKARI\/} Imaging}

The site was observed as part of the all-sky survey scanning mode of the {\sl AKARI\/} mission with the Far IR Surveyor instrument during 2006 April--2007 August at 65, 90, 140, and 160 $\mu$m.

\subsection{{\sl WISE}/{\sl NEOWISE} Imaging}

The SN site was observed as part of the all-sky survey by the {\sl Wide-Field IR Survey Explorer\/} ({\sl WISE}) at 3.4 (W1), 4.6 (W2), 12 (W3), and 22 $\mu$m (W4) during the cryogenic mission, from 2009 December to 2010 August; by the 3-band mission segment (with W4 no longer useful) from 2010 August to 2010 September; and by the post-cryogenic {\sl NEOWISE\/} from 2010 September to 2011 February, and {\sl NEOWISE}-Reactivation ({\sl NEOWISER}) from 2013 December to the present, in the two shortest-wavelength bands.

\subsection{Gemini Spectroscopy and Imaging}

\subsubsection{GMOS Spectroscopy}

Long-slit spectra were pointedly obtained of SN~2023ixf with GMOS-N \citep{Hook2004} at Gemini-North on 2023 June 3 (MJD 60098.4) as part of program GN-2023A-DD-105 (``Back with a Bang: Gemini Multiwavelength Observations of SN 2023ixf''; PI J.~Lotz).  Observations were taken at the parallactic angle \citep{Filippenko1982} with the $0{\farcs}75$-wide slit using $2 {\times} 2$ binning. The B480 grating was used with two central wavelengths of 5500 and 5600~\AA, chosen to mitigate against chip-gap effects, with $2 {\times} 120$~s exposures at each dither position. A spectrophotometric standard star was observed immediately after the science exposures at a similar airmass.

Raw data were made public immediately via the Gemini Science Archive, and we have reduced them using the {\tt DRAGONS} (Data Reduction for Astronomy from Gemini Observatory North and South) reduction package \citep{Labrie2019}, using the recipe for GMOS long-slit reductions. This includes bias correction, flatfielding, wavelength calibration, and flux calibration.

\subsubsection{'Alopeke Imaging}

SN~2023ixf was also observed on 2023 June 9 with 'Alopeke on Gemini-North, as part of the above DD program. Raw data were made public immediately via the Gemini Science Archive. 'Alopeke \citep{Scott2018,Scott2021} is a resident visiting instrument mounted on the GCAL port of Gemini-North.  A dichroic that splits the incoming light at 6740~\AA\ allows 'Alopeke to obtain simultaneous blue and red images using two identical Andor 1K frame-transfer EMCCDs.  Selectable plate scales support either speckle ($\sim 0{\farcs}001$ pixel$^{-1}$) or wide-field (WF; $\sim 0{\farcs}073$ pixel$^{-1}$) imaging, and SN 2023ixf was observed with both modes.

The speckle observations spanned times 7:54--8:15 (airmass 1.22--1.24) and included 20 sets of $1000 \times 60$~ms exposures, using the 5620 and 8320~\AA\ narrow-band filters (540 and 400~\AA\ wide, respectively), with an EMGAIN of 1000 and a $256 \times 256$ region-of-interest (ROI), yielding a $\sim 2{\farcs}5$ field-of-view (FOV).  These observations were immediately followed by an observation of the PSF standard star HR~5345 (airmass 1.21), consisting of three sets of $1000 \times 60$~ms exposures with an EMGAIN of 20 and 30 in the blue and red, respectively. Data were reduced and final data products produced as described by \citet{Howell2011}.

The 'Alopeke WF observations of the SN spanned 8:34--9:05 and were obtained through Sloan Digital Sky Survey (SDSS) $g'$ and $i'$ filters. To avoid saturating the SN, 3600 exposures of 0.5~s each were collected using an unbinned $768
\times 768$ ROI with an EMGAIN of 5. Bias frames were collected immediately afterward, and twilight flats were obtained in the morning.  All 3600 science exposures were averaged together, bias-subtracted, and flat-fielded to produce the image mosaic we show in Figure \ref{fig:progID}. Both the raw and fully reduced data for these observations are available in the Gemini archive.  

\subsection{Keck Spectroscopy}

SN 2023ixf was observed on 2023 June 7 with the High-Resolution Echelle Spectrometer (HIRES; \citealt{Vogt1994}) on the 10~m Keck I telescope at the W. M. Keck Observatory. Two spectra were obtained using the C2 Decker with dimensions $0{\farcs}86 \times 14\arcsec$ which achieves a resolution $R \approx 60,000$ following the standard setup of the California Planet Search \citep{howard10}. For both spectra, an exposure meter was used to obtain a signal-to-noise ratio (SNR) of 95 per reduced pixel on blaze near 5500~\AA. The first exposure began at 7:31 and lasted 199~s, the second exposure began at 7:35 and lasted 192~s. The spectra were converted into a one-dimensional (1D) format with a wavelength solution according to the methodology described by \citet{Petigura17}. The nominal wavelength solution is accurate to at least 1 HIRES pixel, or $\sim 1$ km s$^{-1}$. We created coadditions of the two exposures for each order.

\subsection{KAIT Photometry}

Follow-up observations of SN~2023ixf were performed by the Katzman Automatic Imaging Telescope (KAIT) as part of the Lick Observatory Supernova Search \citep[LOSS;][]{Filippenko2001}. Multiband $BVRI$ images and additional ``clear'' (unfiltered, close to the $R$ band in response; see \citealt{Li2003}) images  were obtained. Here we  only focus on the $B$ and $V$ data. The full light-curve dataset will be presented elsewhere (W.~Zheng et al., in prep.).

All images were reduced using a custom pipeline\footnote{https://github.com/benstahl92/LOSSPhotPypeline} detailed by \citet[][]{Stahl2019}. PSF photometry was obtained using {\tt DAOPHOT} \citep[][]{Stetson1987} from the {\tt IDL} Astronomy User's Library\footnote{http://idlastro.gsfc.nasa.gov/}. Owing to the small FOV of our images, only one reference star was available for calibration, namely star ``m'' from \citet[][ see their Figure 1]{Henden2012}, with the Landolt magnitudes for the star transformed to the KAIT natural system. Apparent magnitudes were all measured in the KAIT4 natural system, and the final results were transformed to the standard system using the local calibrator and color terms for KAIT4 \citep[see][]{Stahl2019}.

\section{Progenitor Candidate Identification}\label{sec:progID}

In Paper I we showed the identification we assigned for the progenitor candidate to a well-detected source in the shortest two {\sl Spitzer\/} IRAC bands, as did \citet{Kilpatricketal2023},  \citet{Jencson2023}, and \citet{Niuetal2023}. This was quite straightforward, since the source was almost exactly at the refined absolute position given for the SN \citep{Itagaki2023}. No other IR source was anywhere proximate to this position. However, to be certain, it is best to consider relative astrometry rather than rely on absolute astrometry, tying all of the observations in which the candidate is detectable to a common frame.

\begin{figure*}[ht!]
\plottwo{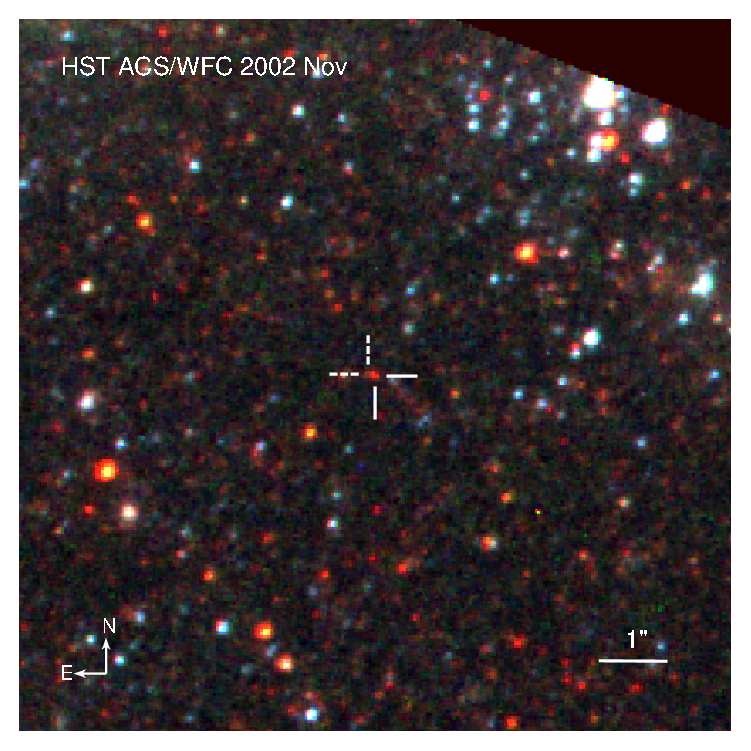}{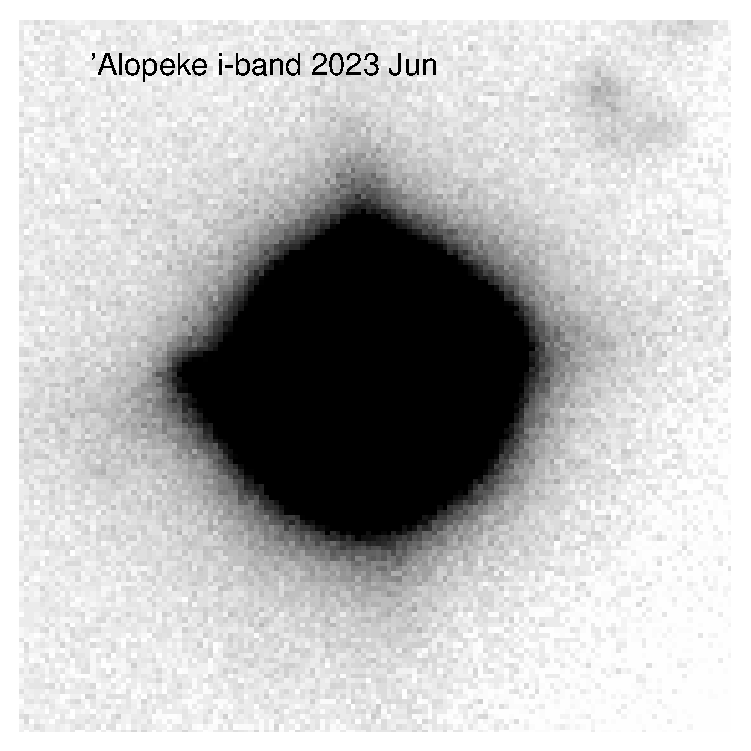}
\caption{{\it Left}: A portion of the {\sl HST\/} ACS/WFC F435W$+$F555W$+$F814W color-composite image mosaic from 2002 November 16, with the progenitor candidate indicated by solid tick marks. (The chip gap can be seen toward the top of the panel.) A putative ``Source B'' is indicated with dashed tick marks (\citealt{Kilpatricketal2023}; see also their figure 1). {\it Right}: A portion of the $i$-band image stack created from Gemini 'Alopeke observations in the Wide Field mode on 2023 June 9, with SN 2023ixf clearly visible. Both panels are shown to the same scale and orientation. North is up and east is to the left.
\label{fig:progID}}
\end{figure*}

We have done this here, first, by astrometrically registering the 'Alopeke WF imaging of the SN from 2023 June to the {\sl HST\/} ACS F814W mosaic from 2002 November. We isolated nine stellar objects in common between the two image mosaics and used the package {\tt photutils.centroids} within {\tt photutils} \citep{photutils} on the nine fiducials in each dataset. We computed average centroids (from the four centroiding methods, {\tt centroid{\textunderscore}com}, {\tt centroid{\textunderscore}1dg}, {\tt centroid{\textunderscore}2dg}, {\tt centroid{\textunderscore}quadratic}) for the fiducials in the 'Alopeke data, with a 1$\sigma$ root-mean-square (rms) uncertainty of 0.70~pixel (50.8~mas, at ${\sim}0{\farcs}0725$~pixel$^{-1}$); and, similarly, for the ACS mosaic, with an rms uncertainty of 0.14~pixel (7.0~mas, at $0{\farcs}05$~pixel$^{-1}$). With the routine {\it geomap\/} within {\tt PyRAF} \citep{Pyraf2012}, with second-order fitting, from the merged fiducial list we were able to align the two mosaics with an rms uncertainty in the mean astrometry of 0.92~ACS/WFC pixel (46.0~mas). We determined the centroid of the SN in 'Alopeke with rms uncertainties of 0.19~pixel (13.9~mas), and transform its position with the routine {\it geoxytran\/} to the F814W mosaic. The overall rms uncertainty in the transformation, adding all of the above uncertainties in quadrature, is estimated to be 70.3 mas, or 1.41 ACS pixel. We isolated the red object we indicate in Figure~\ref{fig:progID} (left panel). 

We measured the centroid of this object with an rms uncertainty of 0.06~ACS pixel (2.8~mas). The SN position, transformed onto the F814W mosaic, differs from the centroid of the candidate by 0.56~ACS pixel (28.1~mas). As can be seen, this difference is within the overall uncertainties in the astrometric transformation. We therefore consider it quite likely that the F814W detection is the progenitor candidate. \citet{Pledger2023}, \citet{Soraisam2023a}, and \citet{Kilpatricketal2023} identified this same object as the progenitor candidate. 

Similarly, we register the Gemini NIRI $K$-continuum band mosaic (see Paper I) to the {\sl HST\/} ACS F814W mosaic, to confirm that the object in NIRI is most likely the same as the one in ACS. We matched 20 stars in common between the two datasets and measured centroids for these with {\tt photutils}, with rms uncertainties of 0.12~pixel (14.3~mas, at $0{\farcs}1171$~pixel$^{-1}$) and 0.09~pixel (4.5~mas) for NIRI and ACS, respectively. The uncertainty in the {\it geomap\/} fit is 0.40 ACS/WFC pixel (20.0 mas), and the uncertainty in the centroid of the object in NIRI is 0.05 pixel (5.6 mas). The total error budget in this registration is then 25.6 mas, or 0.51 ACS pixel. The difference in the object's position transformed to ACS is 0.29 pixel, for a total uncertainty (including that in the F814W centroid) in the transformation of 15.0 mas. The uncertainty in the transformed position is within the total registration uncertainty, and therefore we conclude it is highly likely that the star in F814W and in NIRI $K$-continuum are one and the same.

We dispensed here with performing a further formal registration of the NIRI data to the {\sl Spitzer\/} detection, since we believe it has been convincingly demonstrated that the object in the near-IR and the mid-IR vary at essentially the same frequency and in phase (e.g., \citealt{Kilpatricketal2023,Jencson2023}; Paper I). We consider the likelihood that two stars in immediate proximity to each other would have such strikingly similar and coordinated behavior across photometric bands and not actually be the same object to be extremely low.

We note that, in most of the available {\sl HST\/} data, the progenitor candidate was not detected (Table~\ref{tab:hst_obs}). For these data we provide upper limits on detection, based on the brightnesses of objects in the vicinity of the candidate's location at the formal $5\sigma$ detection level from {\tt Dolphot} (see \citealt{VanDyk2023a} for the rationale for this approach).

\begin{figure}[ht!]
\plottwo{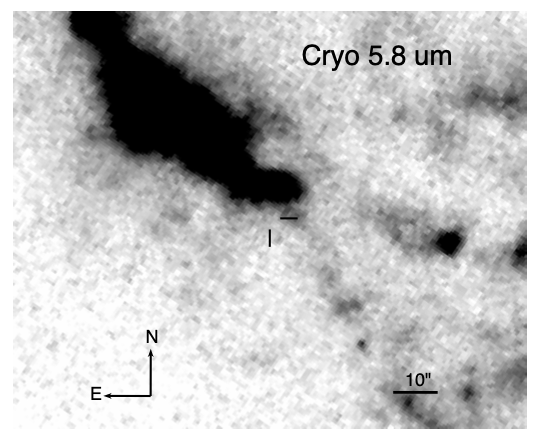}{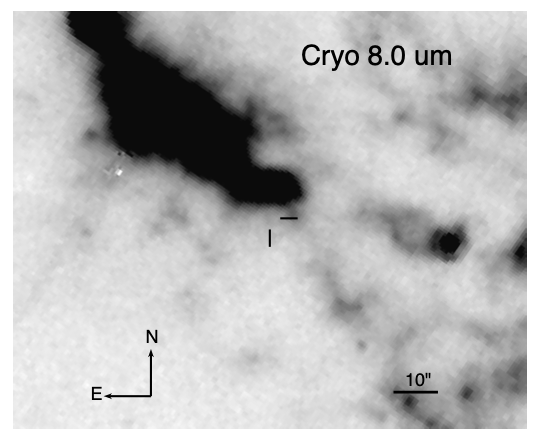}
\caption{{\it Left}: A portion of the {\sl Spitzer\/} IRAC 5.8 $\mu$m image mosaic from 2004, with the SN progenitor candidate location (based on the absolute position) indicated by tick marks. The candidate is not detected in these data. {\it Right}: Same as the left panel, but at 8.0 $\mu$m. Both panels are shown to the same scale and orientation. North is up and east is to the left.
\label{fig:spitzer}}
\end{figure}

The star was also not detected in the 2004 {\sl Spitzer\/} 5.8 and 8.0 $\mu$m data; see Figure~\ref{fig:spitzer}. We processed the data in these two bands in a similar fashion to that described in Paper I, although instead of using {\tt APEX} User List Multiframe \citep{Makovoz2005b}, we simply used {\tt APEX} Multiframe, in the SNR image input mode, to perform point response function (PRF) fitting to individual sources in the data. We located the faintest sources that were detected in the overall environment of the progenitor candidate and, from the brightnesses of these sources, we set upper limits on the candidate's detection of 45 and 128 $\mu$Jy at 5.8 and 8.0 $\mu$, respectively (these limits correspond to SNR$=12$ and 29 in each of these two respective bands).

\begin{figure}[ht!]
\plotone{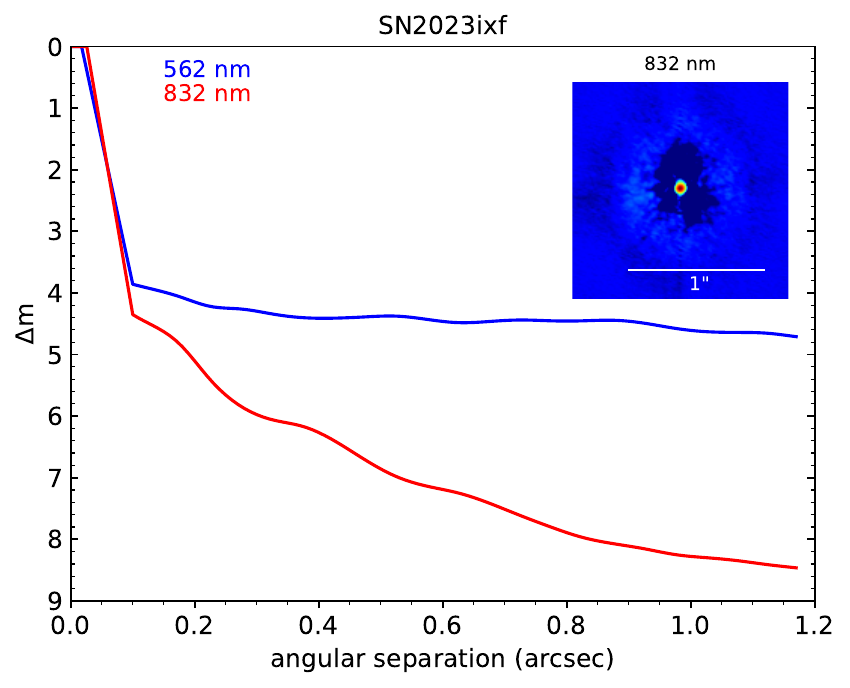}
\caption{Reconstructed speckle image (inset) and differential magnitude detection limits $\Delta m$ of SN 2023ixf obtained with the 'Alopeke speckle camera at Gemini-N on 2023 June 9. From the 8320~\AA\  narrow-band filter observations, we can rule out any neighboring object to $\sim 5$ mag and $\sim 9$ mag below the SN brightness at $0{\farcs}1$ and $1{\farcs}2$, i.e., to $I \approx 16$ and $\approx 20$ mag, respectively.
\label{fig:speckle}}
\end{figure}

Owing to its close proximity to NGC 5461, and the comparatively inadequate spatial resolution and sensitivity of the {\sl Spitzer\/} MIPS, {\sl Herschel\/} and {\sl AKARI\/} instruments, and (to a lesser extent) {\sl WISE\/} W3 and W4, the progenitor candidate was completely overwhelmed by the luminous emission from that giant H~{\sc ii} region at wavelengths 
$\geq$ 11 $\mu$m. We estimated the flux at the exact SN position in these datasets and assumed these fluxes as the upper limits on the progenitor's brightness. None of these limits were particularly constraining: respectively, $<0.025$, $<0.64$, and $3.43$ Jy at {\sl Spitzer\/} 24, 70, and 160 $\mu$m; $<0.018$ and $<0.252$ Jy at {\sl Herschel\/} PACS 70 and 160 $\mu$m; $<0.296$, $<0.161$, and $<0.058$ Jy at {\sl Herschel\/} SPIRE 250, 350, and 500 $\mu$m; and, $<3.78$, $<3.29$, $<6.37$, and $<1.17$ Jy at {\sl AKARI\/} 65, 90, 140, and 160 $\mu$m. For {\sl WISE\/} W3 and W4 we estimated limits that were considerably more constraining: $<0.00031$ and $<0.00125$ Jy, respectively.

We did visually inspect all of the hundreds of publicly-available {\sl WISE\/} and {\sl NEOWISE\/} W1 and W2 single exposures up to 2022 May 24, in which the site is distinguishable from its general environment, and the progenitor candidate was not detectable in any of these frames \citep[see also][]{Hiramatsu2023}. 
Based on the 3$\sigma$ detections of stars within 60\arcsec\ of the SN position in each of these datasets, we estimated respective limits on the progenitor candidate detection of $>15.83$ and $>16.33$ mag ($<0.000144$ and $<0.000050$ Jy) for cryogenic W1 and W2; $>16.24$ and $>15.14$ mag ($<0.000099$ and $<0.000151$ Jy) for post-cryogenic W1 and W2; and, $>16.42$ and $>14.88$ mag ($<0.000084$ and $<0.000192$ Jy) for {\sl NEOWISE\/} W1 and W2. That the star was not detected is not surprising, since we found in Paper I that the {\em peak\/} brightnesses were $\sim 17.0$ and $\sim 16.5$ mag in {\sl Spitzer\/} 3.6 and 4.5 $\mu$m (respectively) between 2004 March and 2019 October. 
These limits do, however, rule out any luminous outbursts at these wavelengths in the 2019 October to 2022 May time period, i.e., $\sim 1300$ to $\sim 360$~d prior to explosion. This is consistent with the findings by \citet{Jencson2023}, based on their $J$- and $K$-band data.

We note that \citet{Kilpatricketal2023} identified a ``Source B'' in the {\sl HST\/} F814W image mosaic, $\sim 0{\farcs}1$ from the progenitor candidate, their ``Source A.'' We also recognize the presence of what appears to be an ``appendage'' to the candidate (\citeauthor{Kilpatricketal2023}~characterized this source as being to the northeast of the candidate, whereas to us, it appears more-or-less due east). However, we call into question here whether this is a real object. First, we visually inspected each of the two frames which comprise the mosaic and find that this other source appears more prominently in one frame than the other. Additionally, although {\tt Dolphot}, when run on bands F435W, F555W, and F814W all at once, detects it as a separate source, its object type is ``4,'' which indicates that the object is ``too sharp'' and likely not a good stellar detection. We do find that, when {\tt Dolphot} is run in F814W and F555W separately (the source is definitely not detectable at F435W), the routine does identify it as a stellar source, albeit too faint for PSF determination in the latter band. If real, we estimate the \citet{Kilpatricketal2023} ``Source B'' is less than a third as bright as the progenitor candidate, so its influence on the photometric measurements of the candidate is relatively minimal, which, as we show in Section~\ref{sec:sedfit}, would have little effect on the characterization of the progenitor's properties, particularly given the likely variability of the candidate at F814W.

Finally, we can constrain the presence of a neighboring source at the SN location, based on the 'Alopeke speckle observations; see Figure~\ref{fig:speckle}.
The speckle image exhibits no indication of any fainter, neighboring source at $0{\farcs}1$ to a level of $\sim 5$ mag below the SN brightness, out to nearly 9 mag at $1{\farcs}2$, at the F814W ($\sim I$-band) equivalent. On June 9, when the speckle observations were obtained, the SN was at $I=11.10$ mag (and $V=10.81$ mag) from the KAIT photometry. Hence, neighbors can be ruled out to $I \approx 16$ mag at $0{\farcs}1$, and $\sim 20$ mag at $1{\farcs}2$. While we cannot from these observations eliminate a much fainter star, in particular, the presumed ``Source B,'' we can say that, at least for the SN in its bright state on the plateau, there would be a negligible effect on the SN light from any neighboring object. 

\section{Host Galaxy Distance Estimation}\label{sec:distance}

A number of modern, reliable distance estimates exist for M101, either from Cepheid or from tip-of-the-red-giant branch (TRGB) measurements. For the latter these include distance moduli $\mu=29.30 \pm 0.12$ mag (distance $d=7.24 \pm 0.40$ Mpc; \citealt{Lee2012}); $29.16 \pm 0.13$ mag ($6.79 \pm 0.41$ Mpc; \citealt{Tikhonov2015}); and, $29.07 \pm 0.06$ mag ($6.52 \pm 0.18$ Mpc; \citealt{Beaton2019}). The Cepheid distances include $28.96 \pm 0.11$ mag ($6.19 \pm 0.31$ Mpc; \citealt{Mager2013}); $29.13 \pm 0.19$ mag ($6.70 \pm 0.59$ Mpc; \citealt{Nataf2015}, which is a reanalysis of the distance estimate by \citealt{Shappee2011} with a larger uncertainty); $29.14 \pm 0.09$ mag ($6.71 \pm 0.28$ Mpc; \citealt{Foley2020}); and, $29.18 \pm 0.04$ mag ($6.85 \pm 0.13$ Mpc; \citealt{Riess2022}). Since the TRGB estimates are sensitive to contamination of halo populations by younger stellar populations, and therefore to field choice and availability, our predilection is to select a Cepheid-based distance, and we choose the \citet{Riess2022} distance; to be somewhat more conservative, however, we adopt an uncertainty of 0.1 mag, which is a weighted mean of the uncertainties in the individual Cepheid estimates (as \citealt{Riess2022} pointed out, the Cepheids studied by \citealt{Mager2013}~included stars with long periods that were not adequately sampled and thus could have larger uncertainties in their derived periods;  \citeauthor{Riess2022}~excluded these long-period Cepheids, and we therefore opted not to include the \citeauthor{Mager2013}~result in the distance we adopt here). Therefore, we adopt $d=6.85 \pm 0.32$ Mpc ($29.18 \pm 0.10$ mag) for the distance to SN 2023ixf. (Since the submission of this paper, \citet{Huang2024} presented a somewhat shorter distance, $6.61 \pm 0.18$ Mpc, $\mu=29.10 \pm 0.06$ mag, based on Mira variables; however, as those authors pointed out, this distance agrees with the Cepheid-based distance to within the uncertainties.)

\begin{figure}[ht!]
\plottwo{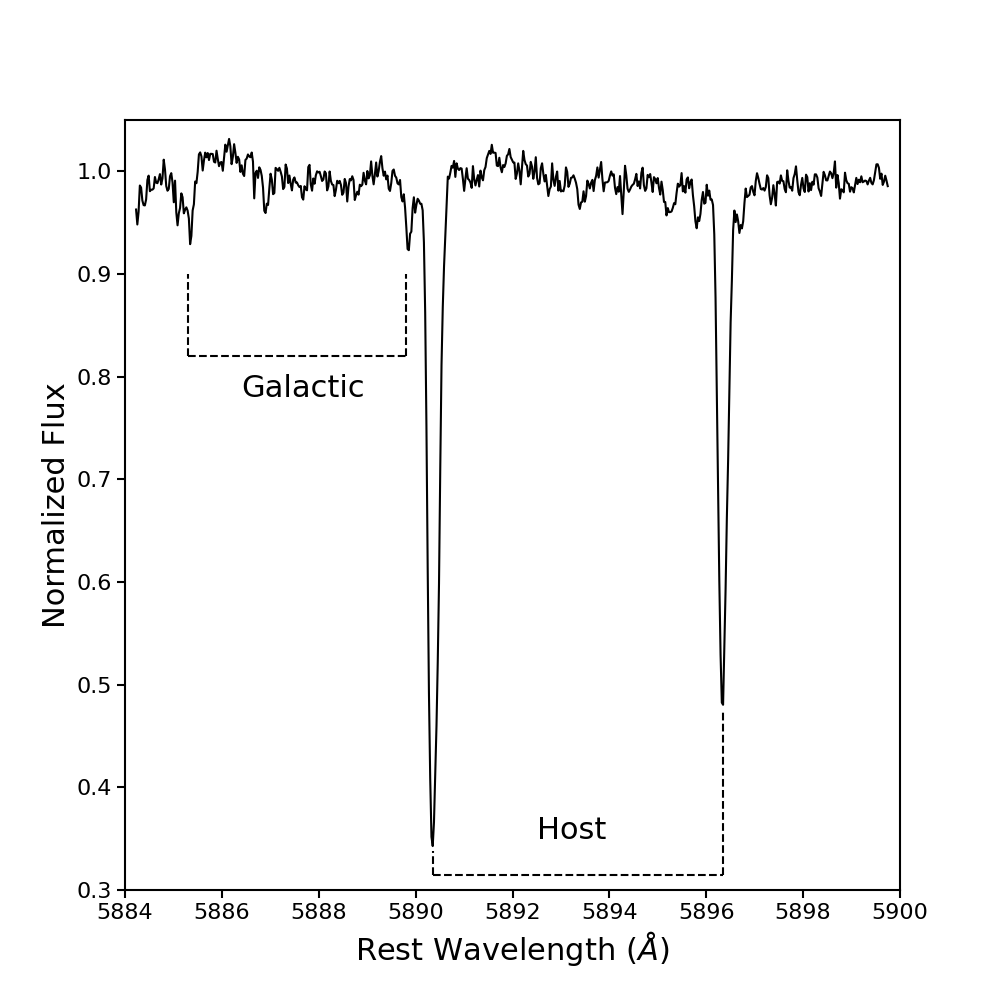}{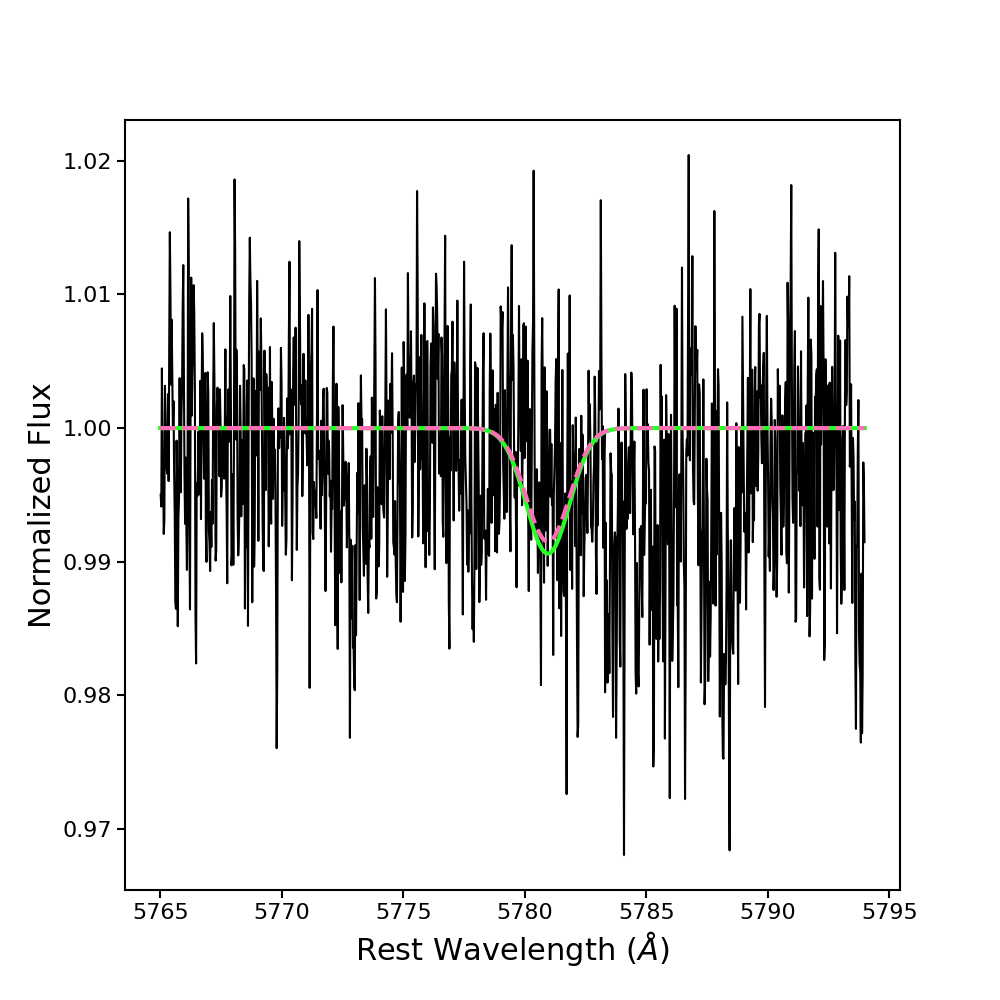}
\caption{{\it Left}: A portion of the Keck HIRES coadded spectrum of SN 2023ixf from 2023 June 7, showing the locations of the Na~{\sc i}~D absorption lines that we attribute to the Galactic foreground and to the host galaxy, as indicated. {\it Right}: A portion of the same spectrum in the region of the DIB $\lambda$5780 feature. We also show a model of the feature, assuming a Gaussian with FWHM $=2.1$~\AA\ and an EQW corresponding to $A_V=0.099$ mag (see \citealt{Phillips2013}), shifted by a velocity offset of $+21$ km s$^{-1}$ (pink dashed curve). Also shown is a similar model, based on a $3\sigma$ upper limit of the EQW (green solid curve). The wavelength axis for both panels is in the host-galaxy rest frame.
\label{fig:HIRES}}
\end{figure}

\section{Host Reddening Estimation}\label{sec:extinction}

Next, it is essential, in order to detail the intrinsic nature of the progenitor candidate, to estimate the amount of extinction to SN 2023ixf. The assumption we have made here is that the total interstellar extinction to the SN is the same as to the progenitor candidate. We are not accounting for any circumstellar extinction related to the candidate; that we will do in Section~\ref{sec:sedfit}. The Galactic foreground contribution to the total extinction is likely to be quite low, $A_V=0.024$ mag (\citealt{Schlafly2011}, via the NASA/IPAC Extragalactic Database, NED). The question then centers on the contribution internal to the host galaxy. Blue and red objects are seen throughout the {\sl HST\/} image mosaic, shown in Figure~\ref{fig:progID}, in the general vicinity of the progenitor candidate --- in fact, a blue star is seen just $0{\farcs}3$ to the southeast of the candidate --- and the reddening appears patchy and variable. The reddening to the progenitor candidate is likely not extreme, since, as can be seen in Figure~\ref{fig:spitzer}, the star appears visually to be located in a region of relatively low (although not negligible) 8~$\mu$m emission --- the IRAC 8.0 $\mu$m band probes the strong molecular complex, composed of the 7.7, 8.3, and 8.6 $\mu$m polycyclic aromatic hydrocarbon features, often associated with interstellar dust \citep[e.g.,][]{Gordon2008}.

We can look to constrain the host reddening from spectroscopic data. This has already been attempted by both \citet{Lundquist2023} and \citet{Smith2023}, based on high-spectral-resolution observations of the SN, effectively using SN 2023ixf as a bright light bulb behind the dusty screen. Both of these studies concluded that the host reddening is relatively low, at $E(B-V)=0.031$ mag, based on the equivalent widths (EQW) of the Na~{\sc i}~D absorption lines. From our Keck HIRES spectrum we performed a similar measurement of the Na~{\sc i}~D EQW; this portion of the spectrum is shown in Figure~\ref{fig:HIRES}. We have identified the features we assign, based on their wavelengths, to both the Galactic component and the host-galaxy component. For the host contribution we measure the Na~{\sc i}~D2 EQW of $0.177 \pm 0.001$~\AA, and for D1, $0.122 \pm 0.003$~\AA. These measurement values are overall quite similar to those that \citet{Lundquist2023} quoted, although the values measured by \citet{Smith2023} are somewhat larger than ours. We found that the two lines were offset relative to the host galaxy redshift by $+21.4$ km s$^{-1}$ (interestingly, \citealt{Smith2023} estimated this offset as $+7 \pm 1$ km s$^{-1}$). Following these two studies, we applied the relations provided by \citet{Poznanski2012} to convert EQW to reddening and found an average (from the three relations, for D1, D2, and D1$+$D2) of $E(B-V)=0.032_{-0.009}^{+0.012}$ mag. Again, overall this is consistent with the estimates made by \citet{Lundquist2023} and \citet{Smith2023}, as well as by \citet{Jacobson-Galan2023}. If we assume that the dust in M101 is similar to Galactic diffuse interstellar dust, then we adopt $R_V=3.1$, for a visual extinction internal to the host of $A_V=0.099_{-0.027}^{+0.037}$ mag.

We also measured the EQW of the lines we associated with the Galactic foreground contribution and found that the D2 EQW $=0.020 \pm 0.004$~\AA\ and the D1 EQW $=0.018 \pm 0.001$~\AA. Again, following \citet{Poznanski2012}, we computed that $E(B-V)=0.016_{-0.004}^{+0.006}$ mag and, therefore, $A_V=0.050_{-0.014}^{+0.019}$ mag. This is about a factor of two larger than the NED value, $A_V=0.024$ mag. Since the overwhelming majority of the detected flux from the progenitor candidate emerges in the IR (see Section~\ref{sec:sedfit}), the effects of reddening turn out to be quite minor. The Galactic foreground component is comparatively small, regardless of whether we adopt the value inferred from the HIRES spectrum or from \citet{Schlafly2011} directly; we choose to adopt the latter for the foreground extinction, thus keeping us consistent with the assumed value in the previous studies of SN 2023ixf.

We also attempted to locate the diffuse interstellar band (DIB) $\lambda$5780 feature in the HIRES spectrum. As \citet{Phillips2013} demonstrated, there appears to be a correlation between the EQW of this DIB feature and $A_V$. We modeled the feature assuming a Gaussian shape with full-width at half-maximum intensity (FWHM) of 2.1 \AA\ and an EQW corresponding to $A_V=0.099$ mag \citep[see][ their equation 6]{Phillips2013}, the internal extinction value we inferred from the Na~{\sc i}~D lines. We also shifted the model in wavelength to match the velocity offset of $+21$ km s$^{-1}$ measured from the Na lines. In addition, we produced a model of the feature based on a $3\sigma$ upper limit on the feature's EQW. As can be seen in Figure~\ref{fig:HIRES}, unfortunately, the HIRES spectrum was not of sufficient SNR for the feature to have been detected.

\begin{figure}[ht!]
\plotone{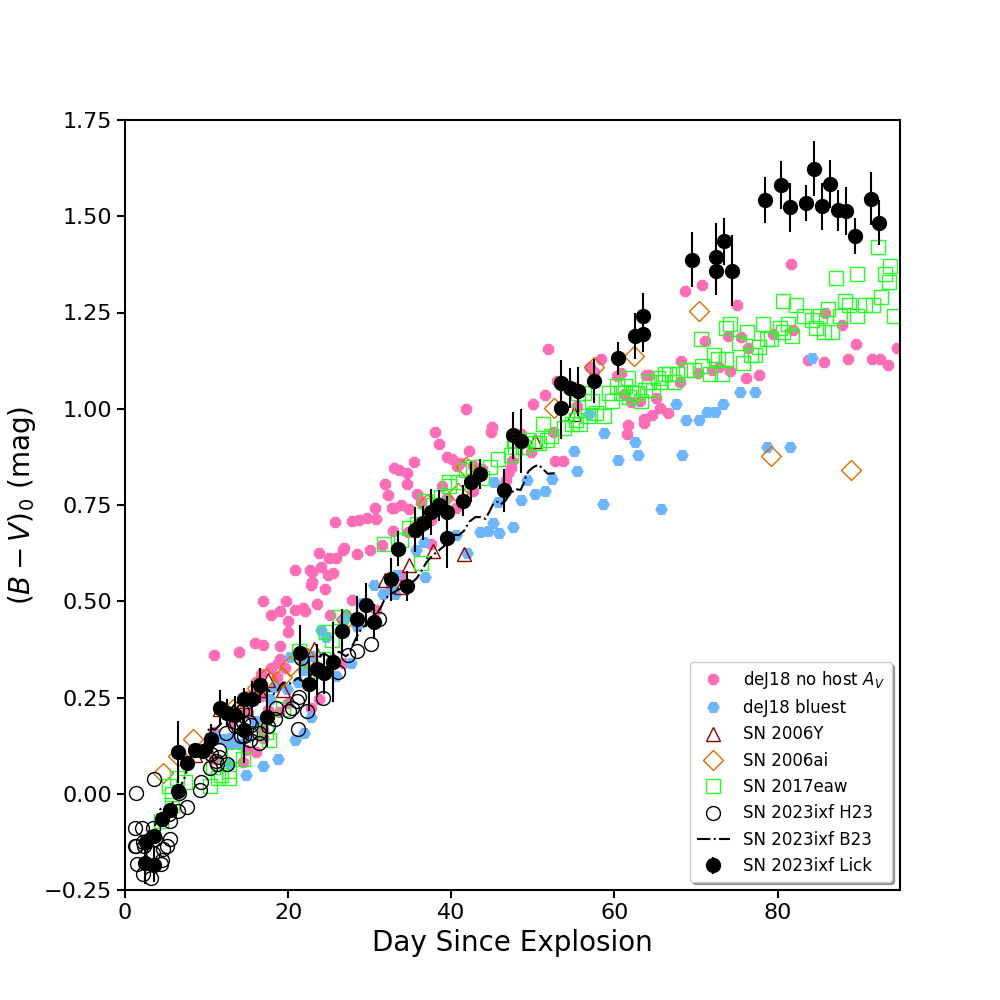}
\caption{$(B-V)_0$ colors for SN 2023ixf from Lick KAIT observations (solid circles), from \citet[][ H23, open circles]{Hiramatsu2023}, and from \citet[][ B23, dot-dashed curve]{Bianciardi2023}. Also shown are the two samples of SNe II-P, one with the bluest colors and the other with no significant host-galaxy reddening, from \citet{deJaeger2018}. Additionally, we show the colors for the SN II-P 2017eaw in NGC~6946 \citep{VanDyk2019}, which likely experienced only Galactic foreground reddening, and two short-plateau SNe, SN 2006Y and SN 2006ai \citep{Hiramatsu2021}, which also experienced minimal host reddening \citep{Anderson2014}. All of the SNe shown, therefore, were dereddened only by the Galactic foreground contribution.
\label{fig:bminv}}
\end{figure}

Finally, we can at least qualitatively assess the effect of any host reddening on SN 2023ixf, based on a color comparison between $B-V$ obtained with KAIT and samples of SNe~II-P with the bluest colors and with no significant host-galaxy reddening from \citet{deJaeger2018}. We also include the luminous SN 2017eaw, for which \citet{VanDyk2019} concluded that the reddening was almost entirely from the Galactic foreground, as well as two short-plateau SNe, SN 2006Y and SN 2006ai \citep{Hiramatsu2021}, which also experienced very little host reddening \citep{Anderson2014}. We show this comparison in Figure~\ref{fig:bminv}. The colors for all of the SNe shown have only been dereddened by removal of the Galactic foreground component in each case. While a fair amount of observational scatter exists in the SN 2023ixf data, the colors of the SN seem to be consistent with the other events with which we compare. \citet{deJaeger2018} cautioned that the physical origins of observed SN~II color differences are not well understood, and factors, such as CSM interaction, possibility of dust destruction and variable reddening with time, further complicate our understanding, such that a significant dispersion in intrinsic SN~II colors precludes the existence and development of a ``color template.'' Nevertheless, any color offset for SN 2023ixf, relative to the comparison samples, appears to be relatively minimal, further implying that the host reddening to the SN must be low.

Note that by day $\sim 57$, the SN 2023ixf color began to diverge redward relative to the other SNe shown in Figure~\ref{fig:bminv}. From more extensive photometric information available for the SN (W.~Zheng et al., in preparation), it is evident that by day $\sim 60$ the SN was already falling off the plateau in both $B$ and $V$. By day $\sim 80$ the SN was approaching the exponential decline tail --- hence, by that age SN 2023ixf had reached a $B-V$ color ($\sim 1.5$ mag) that, for example, the more ``photometrically-normal'' Type II-P SN 2017eaw did not reach until day $\sim 120$ \citep[e.g.,][]{VanDyk2019}. Therefore, SN 2023ixf was simply experiencing the expected, normal color evolution, only ``accelerated'' as a result of the comparatively shorter plateau. We are unsure what impact this had on the circumstellar environment --- only that the progenitor likely had less envelope mass than, for instance, the SN 2017eaw progenitor, and hence the H recombination wave passed through the cooling ejected envelope over a shorter timescale. The lower envelope mass is entirely consistent with the existence of extensive CSM around the progenitor star.

The total, Galactic plus internal host, extinction we adopt hereafter for SN 2023ixf therefore is $A_V=0.12$ mag. We further adopt the \citet{CCM89} reddening law in the optical and near-IR, and \citet{Indebetouw2005} for the mid-IR, with $R_V=3.1$ (appropriate for diffuse interstellar dust). Ultimately, however, it matters little which laws we adopt here, since the total reddening is relatively low.

\section{SN Site Metallicity}\label{sec:metallicity}

At the adopted distance of 6.85 Mpc, SN~2023ixf is located $\sim 8.8$~kpc from the center of M101. From the radial gradient in the oxygen abundance (often assumed as a proxy for metallicity) established by \citet{GarnerM10122}, $12+\log[{\rm O/H}]= (9.21 \pm 0.01) + (-0.039 \pm 0.001) \times R$ dex~kpc$^{-1}$, we computed an abundance value of $8.86 \pm 0.01$.  While this is that study's preferred abundance gradient, based on the calibration from \citet[][ R.~Garner, priv.~comm.]{KK04}, we also obtained values of $8.74 \pm 0.01$ and $8.42 \pm 0.01$ based on the gradients calibrated using \citet{M91} and \citet{PT05}, respectively; see \citet{GarnerM10122}, their table 5. This latter value is consistent with $12+\log({\rm O/H}) = 8.41$ that \citet{Esteban2020} estimated for the neighboring giant H~{\sc ii} region NGC~5461.

\begin{figure}[ht!]
\plotone{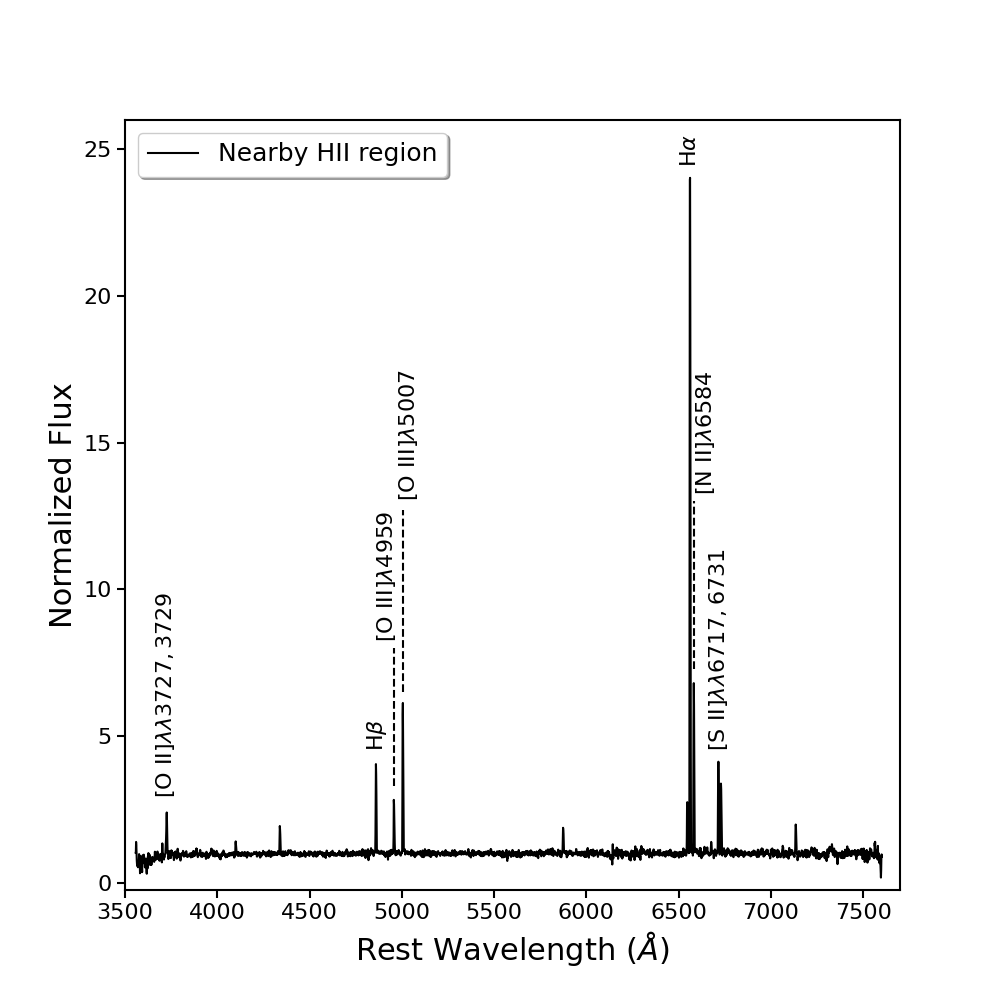}
\caption{Gemini GMOS-N spectrum of emission from the outskirts of H~{\sc ii} region \#1086 \citep[H1086;][]{Hodge1990} nearby ($\sim 8{\farcs}3$) to SN 2023ixf. See Figure~\ref{fig:hiiregion_map}. The spectral continuum has been normalized, however the spectrum as shown has not been reddening-corrected. The locations of the various strong lines we measured, in order to estimate the oxygen abundance from the emission site, are indicated.
\label{fig:hiiregion}}
\end{figure}

To attempt to break this discrepancy, we measured the line strengths from a spectral extraction of nearby emission at the outskirts of H~{\sc ii} region \#1086 (H1086), as catalogued by \citet[][ the center of that region is $\sim 8{\farcs}3$, or $\approx 274$~pc, northwest of SN~2023ixf]{Hodge1990}, that fell on the slit of the Gemini-N/GMOS observations obtained on 2023 June 3; see Figure~\ref{fig:hiiregion}. \citet{Pledger2023} also noted that this H~{\sc ii} region was in proximity of the SN. The extraction of the emission was accomplished using the trace of the SN across the detector as a guide. To deredden the continuum-normalized spectrum, we measured the Balmer decrement (H$\alpha$/H$\beta$ = 10.10), assumed Case B recombination ratios (H$\alpha$/H$\beta$ = 2.86; \citealt{Osterbrock1989}), and then applied the \cite{CCM89} extinction law. This resulted in $E(B-V) = 1.08$ mag. We then measured the EQW of the various strong lines from the dereddened spectrum.

\begin{figure}[ht!]
\plotone{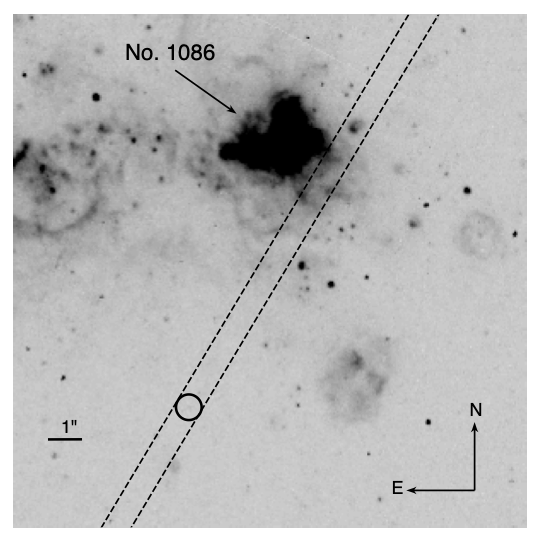}
\caption{Outline of the Gemini GMOS-N $0{\farcs}75$ slit (dashed lines) for the 2023 June 3 spectral observations, overlaid on a portion of the {\sl HST\/} ACS/WFC F658N image mosaic from 2004 February 10. The position of SN 2023ixf is indicated with a circle. Along with the SN, the slit position intersected the outskirts of H~{\sc ii} region \#1086 \citep[H1086;][]{Hodge1990}, the center of which is $\sim 8{\farcs}3$ northwest of the SN.
\label{fig:hiiregion_map}}
\end{figure}

We compared our measurements with those by \citet{KennicuttGarnett96}, who had spectroscopically observed this H~{\sc ii} region previously. Any differences we found may arise from the fact that \citet{KennicuttGarnett96} presumably centered on the H$\alpha$ peak of H1086, whereas, as we show in Figure \ref{fig:hiiregion_map}, the GMOS slit passed through the southwestern edge of that H~{\sc ii} region. In the first place, the reddening value we estimated is larger than that for H1086 from \citet{KennicuttGarnett96}; however, it is reasonable to assume that the reddening may well be variable across the region. Furthermore, \citet{KennicuttGarnett96} measured higher EQW for [O~{\sc ii}] and [O~{\sc iii}], whereas our EQWs for [S~{\sc ii}] are higher by a factor of $\sim 1.5$, while our EQWs for [N~{\sc ii}] are comparable to those of \citeauthor{KennicuttGarnett96}. We measured an EQW for H$\beta$ of 185~\AA, while the \citet{KennicuttGarnett96} value is 292~\AA. We analyzed the various line ratios from our measurements using the strong-line diagnostics from \citet{Curti2020}. The resulting individual abundance values are shown in Table \ref{tab:metallicity}. Note that \citet{KennicuttGarnett96} estimated the value of the R$_{23}$ indicator as 5.02, whereas we computed it to be 3.63. Similarly, following the \citet{Curti2020} calibration with the \citet{KennicuttGarnett96} value, the O abundance of the H1086 region would be $12+\log({\rm O/H}) = 8.56$, whereas we found 8.67 for the emission closer to the edge.

In short, we estimated from the Gemini spectrum a range of values between $8.33 \lesssim [12+\log({\rm O/H})] \lesssim 8.93$, depending on the diagnostic (and including the uncertainties), mirroring the same spread in O~abundance found by \citet{GarnerM10122} and \citet{Esteban2020}. Assuming the most recent value for the solar O~abundance of $12+\log({\rm O/H}) = 8.77 \pm 0.04$ (\citealt{Magg2022}; note that those authors justified their discrepancy with the $8.69 \pm 0.04$ from \citealt{Asplund2021}), the range of O abundances from the nearby emission tends to imply --- assuming that emission at that location is representative of the SN~2023ixf site --- that the metallicity at the SN site is somewhat subsolar to somewhat supersolar. The O$_{3}$N$_{2}$ indicator, in particular, one of the more robust (and nondegenerate) metallicity diagnostics considered here \citep[e.g.,][]{Kewley2008}, tends to point toward above solar. We will therefore consider solar, subsolar, and supersolar metallicities below when analyzing the properties of the progenitor candidate.

We note that \citet{Zimmerman2024} inferred from absorption lines detected in {\sl HST\/} ultraviolet (UV) spectra of the SN, likely arising from the star's CSM,  that the CSM metallicity was approximately solar. From observations of emission regions near the star those authors also inferred solar to slightly subsolar metallicity. Moreover, we note that any post-SN spectroscopic analyses of the host site's metallicity will likely be at least somewhat contaminated by the light of the SN itself for several years to come.

\begin{deluxetable*}{cccc}
\tablewidth{0pt}
\tablecolumns{4}
\tablecaption{Strong-Line Metallicity Diagnostics\label{tab:metallicity}}
\tablehead{\colhead{Indicator} & \colhead{Line Ratio} & \colhead{Value}&\colhead{$12+\log[{\rm O/H}]$}}
\startdata
       R$_{3}$ & [O~{\sc iii}]$\lambda\lambda$4959,5007/H$\beta$ & 1.41 & $8.55 \pm 0.07$ \\
       N$_{2}$ & [N~{\sc ii}]$\lambda$6584/H$\alpha$ & 0.23 & $8.61 \pm 0.10$ \\
       S$_{2}$ & [S~{\sc ii}]$\lambda\lambda$6717,6731/H$\alpha$ & 0.22 & $8.43 \pm 0.10$ \\
       R$_{23}$ & ([O~{\sc ii}]$\lambda$$\lambda$3727,3729+[O~{\sc iii}]$\lambda\lambda$4959,5007)/H$\beta$ & 3.63 & $8.67 \pm 0.12$ \\
       O$_{3}$O$_{2}$ & ([O~{\sc iii}]$\lambda$5007)/([O~{\sc ii}]$\lambda\lambda$3727,3729) & 0.81 & $8.47 \pm 0.14$ \\
       RS$_{32}$ & ([O~{\sc iii}]$\lambda$5007/H$\beta$)+([S~{\sc ii}]$\lambda\lambda$6717,6731/H$\alpha$) & 1.62 & $8.58 \pm 0.12$ \\
       O$_{3}$S$_{2}$ & ([O~{\sc iii}]$\lambda$5007/H$\beta$)/([S~{\sc ii}]$\lambda\lambda$6717,6731/H$\alpha$) & 6.46 & $8.53 \pm 0.11$ \\
       O$_{3}$N$_{2}$ & ([O~{\sc iii}]$\lambda$5007/H$\beta$)/([N~{\sc ii}]$\lambda$6584/H$\alpha$) & 0.32 & $8.84 \pm 0.09$ \\
\enddata
\tablecomments{Definitions of line ratios and values derived for the indicator calibrations are from \citet{Curti2020}.}
\end{deluxetable*}

\section{SED Fitting}\label{sec:sedfit}

We assembled all of the data we have collected for the progenitor candidate, from the optical to the mid-IR, over a span of years from 1999 through 2019. We demonstrated in Paper I that the progenitor candidate is highly variable in the IR (see also \citealt{Kilpatricketal2023,Jencson2023,Niuetal2023}). Notable gaps exist in the coverage of its light curve, so we cannot strictly rule out the existence of excursions beyond regular variability. However, we have shown in Paper I that the observed data are consistent with periodic behavior. In order to model the observed SED we first established the mean brightness in each of the observed bands. For $JHK_s$, this could be accomplished in two different ways, either by simply averaging the observed data or by using the reconstructed light curves (making sure to restrict this to some integer factor of the $\sim 1091$~d period). For the former, the result is $J=20.39$, $H=19.63$, and $K_s=18.61$ mag. For the latter, the values are quite similar: $J=20.39$, $H=19.61$, and $K_s=18.60$ mag. Uncertainties ($1\sigma$) from the observational data, without any weighting, are 0.07, 0.06, and 0.04 mag in $J$, $H$, and $K_s$, respectively.

For the {\sl Spitzer\/} data, we combined all of the Warm (non-cryogenic) data in each band, effectively averaging over all of those observations --- note that the single cryogenic data point in each band was excluded here. At 3.6 $\mu$m, 268 individual Warm {\sl Spitzer\/} frames contain the progenitor site. We ran {\tt MOPEX} \citep{Makovoz2005a} and {\tt APEX} Multiframe on those frames, following the procedures in Paper I, and extracted a flux density for the progenitor candidate of $31.12 \pm 0.12$ $\mu$Jy at SNR = 252 (at the position $\alpha = 14^{\rm h}03^{\rm m}38{\fs}572$, $\delta = +54\arcdeg 18\arcmin 42{\farcs}11$; J2000). Similarly, there are 256 frames at 4.5 $\mu$m. From that band we extracted a flux density of $25.18 \pm 0.09$ $\mu$Jy at SNR = 293 (position $\alpha = 14^{\rm h}03^{\rm m}38{\fs}587$, $\delta = +54\arcdeg 18\arcmin 42{\farcs}06$). Note that these flux densities have been aperture- and pixel-phase-corrected\footnote{See https://irsa.ipac.caltech.edu/data/SPITZER/\newline{docs/irac/iracinstrumenthandbook/57/\#{\textunderscore}Toc82083747.}}; however, they were not color-corrected\footnote{See https://irsa.ipac.caltech.edu/data/SPITZER/\newline{docs/irac/iracinstrumenthandbook/15/\#{\textunderscore}Toc82083614.}} (although such correction for a $\sim 3000$~K blackbody is effectively unity).

To convert the {\sl Spitzer\/} flux densities into Vega magnitudes, we adopted the IRAC zeropoints, $272.2\pm 4.1$ and $178.7\pm 2.6$ Jy at the nominal channel wavelengths of 3.544 and 4.487 $\mu$m, respectively. We list the measured magnitudes (Vega) in all of the bands in Table~\ref{tab:progenitor}. We also provide in the table both the assumed Galactic foreground and internal host reddening in each band, as well as the resulting reddening-corrected flux densities.

Since the original submission of this paper, \citet{Liu2023} published a detection of the SN progenitor candidate in the SDSS $z$ band at $22.78 \pm 0.06$ mag, and we have now included this in our analysis.

\begin{deluxetable*}{cccccccc}
\tablewidth{0pt}
\tablecolumns{8}
\tablecaption{Adopted Brightnesses of the Progenitor Candidate \label{tab:progenitor}}
\tablehead{\colhead{Band} & \colhead{$\lambda_{\rm eff}$} & \colhead{$m_{\rm obs}$} & \colhead{$A_{\lambda}$(Gal)} & \colhead{$A_{\lambda}$(host)} & \colhead{Zero Point} & \colhead{$f^0_{\nu}$} & \colhead{$\sigma{f^0_{\nu}}\ (+,-)$}\\
\colhead{} & \colhead{($\mu$m)} & \colhead{(Vegamag)} & \colhead{(Vegamag)} & \colhead{(Vegamag)} & \colhead{(Jy)} & \colhead{($\mu$Jy)}  & \colhead{($\mu$Jy)}}
\startdata
F658N      & 0.6586 & 24.53(0.17) & 0.020 & 0.081(0.026) & 2569.107 &  0.434 & 0.075, 0.064 \\
F675W      & 0.6768 & 24.47(0.20) & 0.019 & 0.078(0.025) & 2913.855 &  0.346 & 0.090, 0.071 \\
F673N      & 0.6888 & 24.91(0.25) & 0.019 & 0.077(0.025) & 2911.375 &  0.518 & 0.106, 0.088 \\
F814W      & 0.8419 & 24.31(0.05) & 0.013 & 0.056(0.018) & 2457.908 &  0.493 & 0.023, 0.022 \\
$z$        & 0.9122 & 22.78(0.06) & 0.011 & 0.047(0.015) & 2238.052 &  1.828 & 0.113, 0.106 \\
$J$        & 1.2485 & 20.39(0.07) & 0.006 & 0.028(0.009) & 1599.972 & 11.524 & 0.774, 0.725 \\
$H$        & 1.6510 & 19.61(0.06) & 0.004 & 0.018(0.006) & 1037.303 & 15.158 & 0.865, 0.819 \\
$K_s$      & 2.1562 & 18.60(0.04) & 0.003 & 0.012(0.004) &  668.271 & 24.592 & 0.927, 0.893 \\
3.6 $\mu$m & 3.5388 & 17.39(0.04) & 0.002 & 0.007(0.002) &  272.200 & 30.476 & 1.145, 1.104 \\
4.5 $\mu$m & 4.4724 & 17.28(0.05) & 0.001 & 0.005(0.002) &  178.700 & 22.033 & 0.975, 0.934 \\
\enddata
\tablecomments{Columns: Observational band; bandpass effective wavelength for a 1761 K blackbody; observed magnitude; Galactic and internal host extinctions at effective wavelength; Vegamag zero point; extinction-corrected flux density; and uncertainty in flux density. Uncertainties in $m_{\rm obs}$ and $A_{\lambda}$(host) are in parentheses.}
\end{deluxetable*}

We incorporated an estimate of the source variability into the flux-density uncertainties as follows \citep[see][ e.g., their section 2.2]{Riebeletal2012}. We adopted the amplitudes estimated in Paper I for the $J$, $H$, $K_{\rm s}$, and {\sl Spitzer\/} 3.6 and 4.5 $\mu$m light curves to estimate the range of flux variation in those bands. \citet[][ see their figure 5a]{Smithetal2002} presented the relationship between the $J$-band and $V$-band amplitudes for their sample of IR-bright Mira variables (which are not too dissimilar from variable RSGs). We fit a line to those data and used it to predict the $V$-band amplitude for the progenitor candidate, from its $J$-band amplitude in Paper I; we obtained a value of $\sim 2.6$ mag, consistent with the source being a long-period variable. This amplitude was used to estimate the range of flux variations for the optical bands, in particular, F814W. We added the corresponding flux variation to the measurement uncertainty in quadrature. We also assumed that the $z$-band variability amplitude was identical to that in the $V$ band.

\begin{figure}[ht!]
\plotone{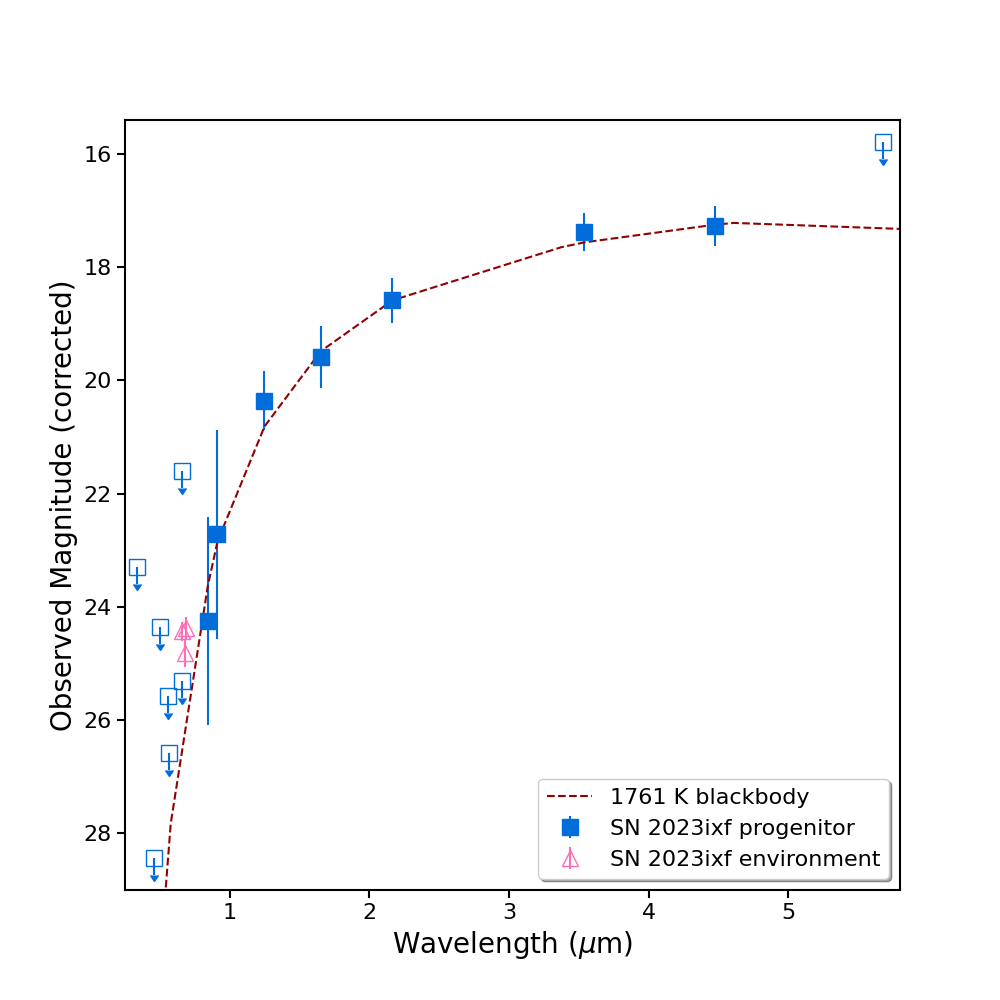}
\caption{The observed SED of the SN 2023ixf progenitor candidate (blue squares), together with detections dominated by emission from the SN environment (magenta triangles; see text), all corrected for both Galactic foreground and host galaxy reddening. The SED can be approximated by a cool blackbody (dashed red curve), of $\sim 1761$~K.
\label{fig:bbsed}}
\end{figure}

We show the reddening-corrected observed SED with the now-inflated uncertainties, as described above, in Figure~\ref{fig:bbsed}. We find that the progenitor candidate SED can be approximated initially by a simple blackbody; an observer within the host galaxy near the star would have assessed it as a rather cool ($\sim 1761$~K) and quite luminous ($\sim 1.11 \times 10^5\ L_{\odot}$, integrating the blackbody and applying our adopted distance to the host) object. The fact that the $z$-band brightness measurement, obtained from an independent investigation, aligns smoothly (both with and without the inflated uncertainties) with our measurements in the other bands provides us with some confidence in the veracity of our methods and our corresponding conclusions.

\begin{figure*}[ht!]
\plotone{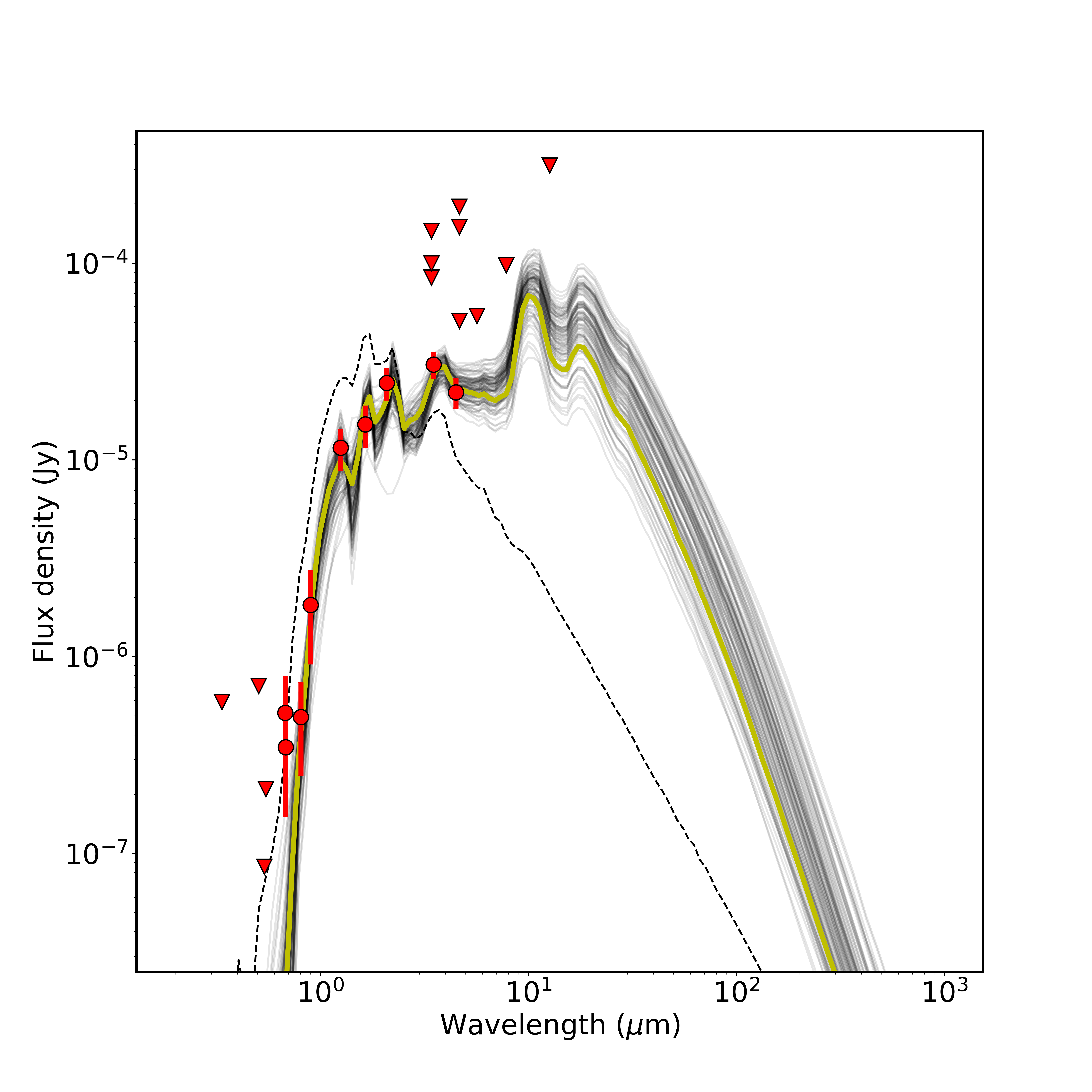}
\caption{The reddening-corrected observed SED (red circles), together with the best-fit {\tt GRAMS} model (solid yellow curve) and posterior samples from the MCMC (gray curves). A {\tt PHOENIX} model (dashed line) at the median estimate for $T_{\rm eff}$, 3000 K, is also shown for comparison. The error bars include a combination of the measurement uncertainties, as well as the range in flux in each band as a result of source variability. Flux upper limits are shown for various {\sl HST\/} bands, the {\sl Spitzer\/} IRAC 5.8 and 8.0 $\mu$m bands, and the {\sl WISE} W1, W2, and W3 bands (red triangles); note that we have shown the three {\sl HST\/} detections shortward of 0.8 $\mu$m as triangles (upper limits) as well. The model predictions are consistent with these limits.
\label{fig:GRAMSfit}}
\end{figure*}

We have performed a more rigorous fit of the SED with O-rich dust models from the \underline{G}rid of \underline{R}ed supergiant and \underline{A}GB \underline{M}odel\underline{S} \citep[{\tt GRAMS};][]{Sargentetal2011,Srinivasanetal2011}; see Figure~\ref{fig:GRAMSfit}. The choice of O-rich dust for the progenitor candidate is justified, given its RSG nature \citep[e.g.,][]{Seab1989}. The {\tt GRAMS} O-rich models are constructed using the {\tt PHOENIX} model photospheres at solar metallicity \citep{Kucinskasetal2005,Kucinskasetal2006} with surface gravity $\log{(g~[{\rm cm~s}^{-2}])}=-0.5$, and with optical constants for O-deficient silicates from \citet{Ossenkopfetal1992}, assuming a spherically symmetric dust shell of inner radius $R_{\rm in}$ (which is effectively set by the condensation temperature of silicate dust) with a constant $\dot M$ (and, hence, an inverse-square density fall-off) and an outer radius 1000 times that of $R_{\rm in}$. (The SN site may be at subsolar or supersolar metallicity; however, even though the input photospheres are at solar metallicity, as we will see the optical depths of the best-fit models are sufficiently high that the underlying stellar characteristics are virtually impossible to distinguish.)

A preliminary fit to the SED with the {\tt GRAMS} models resulted in an optical depth $\tau$ of 1 at 10 $\mu$m; however, the optical depth sampling in this range is quite poor (available models were computed with $\tau=0.5$, 1, and 2), resulting in unreliable estimates of the effective temperature (the best-fit model corresponds to the lowest temperature available in the grid, consistent with the $T_{\rm eff}$ distribution experiencing a slower falloff toward cooler temperatures). Robust parameter estimates derived from fitting the progenitor candidate SED therefore cannot be obtained unless we increase this sampling. We accomplished this by training an artificial neural network on the {\tt GRAMS} grid to predict spectra and dust-production rates (DPRs) for arbitrary parameter combinations. We then performed a Markov Chain Monte Carlo (MCMC) procedure using the \texttt{emcee} package \citep{Foreman-Mackeyetal2013}, employing the neural network to predict spectra for the parameter combinations explored by the MCMC. The MCMC fitting does not fit any of the upper limits; however, the best-fit models are consistent with these limits. We treated all of the other available data as upper limits (including the detections at {\sl HST\/} F656N, F673N, and F675W) and similarly did not fit them.

Another aspect that concerned us was the value of the $K_s$ brightness measured from the Gemini NIRI image. In Paper I we photometrically calibrated this brightness to 2MASS \citep{Skrutskie2006}, as did other studies \citep{Kilpatricketal2023,Jencson2023,Qin2023}. However, the filter through which the NIRI observations were made was not $K_s$; instead, it was a narrower, contiguous $K$-continuum filter used as an ``off-band'' for Br$\gamma$ imaging. (The $K$-continuum bandpass, however, is within the broader $K_s$ bandpass.) For a reddened 1761~K blackbody (referring to Figure~\ref{fig:bbsed}), a difference of $\sim 0.09$ mag exists between the synthetic photometry through these two $K$ filters. Additionally, applying synthetic photometry to the suite of {\tt GRAMS} models, for $J-K_s \approx 3.2$ mag, the approximate color of the progenitor candidate, the photometry differs between the two filters by $0.31 \pm 0.14$ mag (molecular absorption features in the input photospheric emission, only moderately diminished by the silicate dust, fall within the $K$-continuum bandpass; hence, the larger difference compared to a continuum-only blackbody). Furthermore, the MCMC can model for the calibration as a nuisance parameter, and from this treatment we found that the uncertainty in the calibration may be $\lesssim 0.40$ mag. We initially computed fits with and without $K_s$, and found very little difference between the results. This is not surprising, since the SED for this dusty star is quite well-sampled, even in the absence of information in $K_s$. Given the considerable uncertainty in the $K_s$ calibration, we chose hereafter to conduct the fitting without this band.

\bibpunct[,]{(}{)}{,}{a}{}{,}

\begin{figure}[ht!]
\plotone{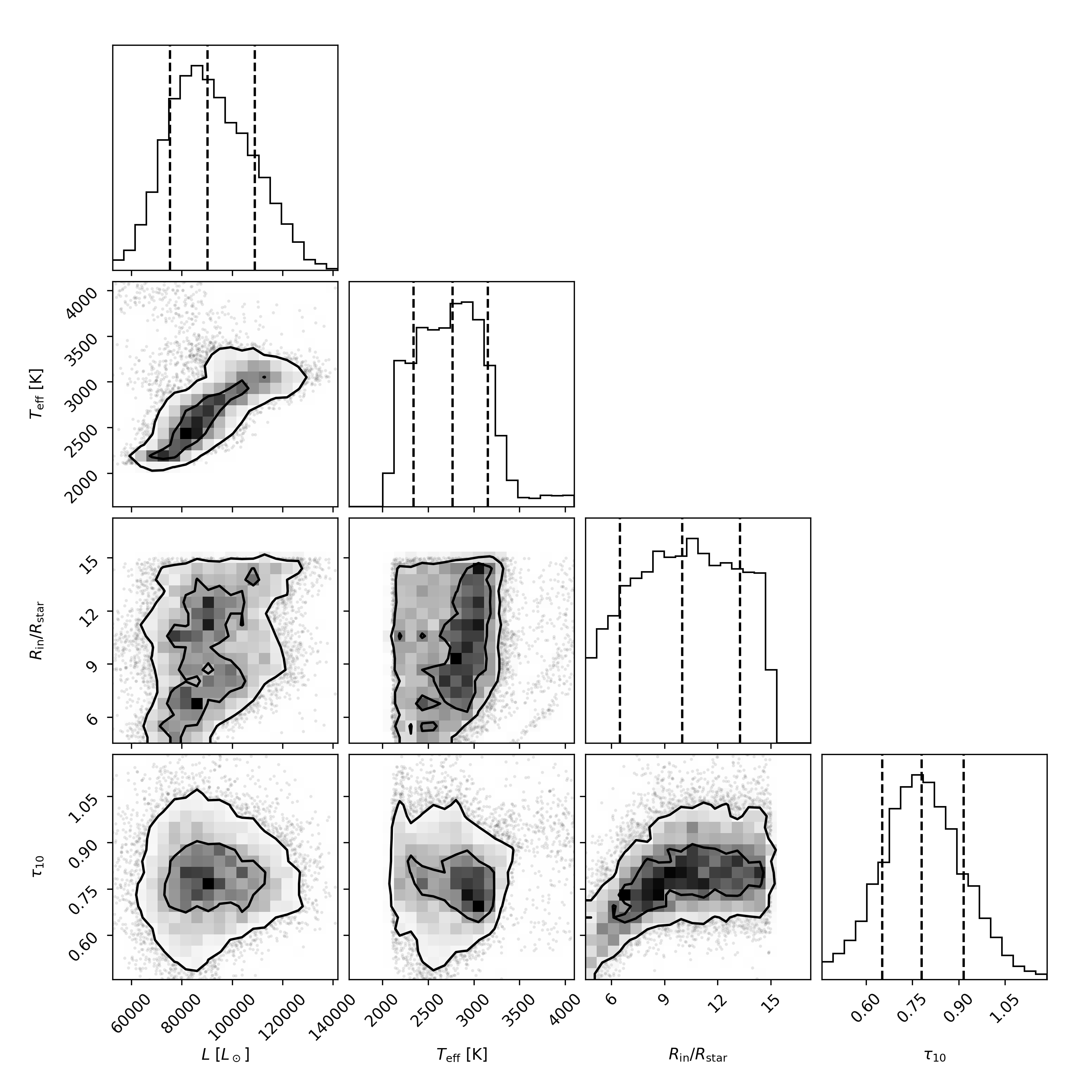}
\caption{Posterior distribution of parameters (luminosity $L_{\rm bol}$, effective temperature $T_{\rm eff}$, inner radius of dust shell $R_{\rm in}$, and 10 $\mu$m optical depth ${\tau}_{10}$) obtained from the MCMC sampling (see text). The scatter plots show the variation of each parameter against every other parameter, with contours containing $\sim 39$\% and 84\% of the probability mass (corresponding to $1\sigma$ and $2\sigma$ contours in 2D). The histograms illustrate the distributions of the individual parameters, with vertical dashed lines representing the 68\% credible interval for each parameter.
\label{fig:MCMCcorner}}
\end{figure}

We present in Figure~\ref{fig:GRAMSfit} the best-fit {\tt GRAMS} model spectrum for the progenitor candidate SED. We show in Figure~\ref{fig:MCMCcorner} a corner plot with the distribution of the resulting values of $L_{\rm bol}$, $T_{\rm eff}$, $R_{\rm in}$, and $\tau$ at 10 $\mu$m (${\tau}_{10}$). While $L_{\rm bol}$ and ${\tau}_{10}$ are well constrained, $T_{\rm eff}$ is not. This is a direct result of the large degree of dust obscuration, which prevents a precise determination of the properties of the photosphere, and the lack of tighter constraints on the dust content owing to the availability of only two mid-IR detections. The range of acceptable $T_{\rm eff}$ values gradually increases with increasing obscuration, as seen in Figure~\ref{fig:MCMCcorner}. This degeneracy is responsible for the large range in predicted $T_{\rm eff}$. The 68\% credible interval for the inner radius of the dust shell is 6.5--13.3 $R_{\star}$. It should be noted that with increasing optical depth, emission from the innermost regions of the shell is obscured enough that the models overestimate the inner radius of the dust shell. Even if this were not the case, the range of values for the inner radius is consistent with the range in which silicate dust is expected to form \citep[densities remain high enough for silicate dust formation out to $\sim 15\ R_{\star}$; see, e.g.,][ their sections 2.1 and 2.2]{Hoefner2007proc}. The corresponding dust density at the inner radius is in the range $10^{-17}$ to $10^{-18}$ g cm$^{-3}$, which is consistent with values expected for the dust-formation zone in cool stars \citep[cf.][ their figure 1; the values in that figure should be scaled down by a factor of 100--200 to convert them to dust densities]{Hoefner2007AA}.

\bibpunct[;]{(}{)}{;}{a}{}{;}

\section{Progenitor Properties}\label{sec:properties}

The best-fit parameters derived from the {\tt GRAMS} grid are given in Table \ref{tab:progenitor_properties}. These parameters describe properties of the photosphere ($L_{\rm bol}$ and $T_{\rm eff}$), as well as the dust shell ($R_{\rm in}$, relative to the effective stellar radius $R_{\star}$; DPR; ${\tau}_{10}$; and, the $\tau$ of the circumstellar shell at 1 $\mu$m, ${\tau}_{1}$). Here, $L_{\rm bol}$ is obtained by integrating under the entire model SED. For each parameter, we show in the table the maximum-likelihood estimate (MLE) from the MCMC sampling, as well as the median (50\%) and 68\% credible interval of values. Based on these results, we find an $L_{\rm bol}$ and DPR of $9.0^{+1.9}_{-1.5}\times 10^4\ L_\odot$ and $6.0^{+6.8}_{-2.0}\times 10^{-8}\ M_\odot$ yr$^{-1}$, respectively, for the progenitor candidate. We quote here the median values, as they are more representative of the typical range of parameter values than the MLE. As discussed in the previous section, $T_{\rm eff}$ is not well constrained; we find a median of 2770~K, with a 68\% credible interval of 2340 to 3150~K. The stellar radius $R_{\star}$ is in the approximate range (921--2012)\ $R_{\odot}$, with median $1389\ R_{\odot}$.

\begin{deluxetable*}{cccccc}
\tablewidth{0pt}
\tablecolumns{6}
\tablecaption{{\tt GRAMS} Best-Fit Parameters for the Progenitor Candidate \label{tab:progenitor_properties}}
\tablehead{\colhead{Parameter} & \colhead{Unit} & \colhead{MLE} & 
\colhead{$p16$} & \colhead{Median} & \colhead{$p84$}}
\startdata
Bolometric luminosity, $L_{\rm bol}$ & $10^4~L_\odot$ & 9.1 & 7.5 & 9.0 & 10.9 \\
Effective temperature, $T_{\rm eff}$ & K & 2710 & 2340 & 2770 & 3150 \\
Optical depth (1 $\mu$m), ${\tau}_{1}$ & \nodata & 1.3 & 1.5 & 1.7 & 2.0 \\
Optical depth (10 $\mu$m), ${\tau}_{10}$ & \nodata & 0.59 & 0.65 & 0.78 & 0.91 \\
Shell inner radius, $R_{\rm in}/R_{\star}$ & \nodata & 5.9 & 6.5 & 10.0 & 13.3 \\
Dust production rate, DPR & $10^{-8}~M_\odot$ yr$^{-1}$ & 3.5 & 4.0 & 6.0 & 12.8 \\
\enddata
\tablecomments{The table lists, for each parameter, the maximum-likelihood estimate (MLE) and the 16$^{\rm th}$, 50$^{\rm th}$, and 84$^{\rm th}$ percentiles ($p16$, median, and $p84$, respectively).}
\end{deluxetable*}

{\tt GRAMS} models assume an expansion velocity of 10 km s$^{-1}$, typical for asymptotic giant branch (AGB) stars and RSGs (the common range observed for RSGs is 10--20 km s$^{-1}$; e.g., \citealt{Decinetal2006,HumphreysJones2022,Decin2024}). \cite{Kilpatricketal2023}, instead, assume 50 km s$^{-1}$, which is more appropriate for hypergiants (e.g., VY~CMa) or for super-winds associated with enhanced mass loss prior to  explosion \citep[e.g.,][]{Shivversetal2015}. Unfortunately, the lack of {\sl Spitzer\/} data near the epoch of explosion makes it challenging to verify directly the super-wind scenario (see Section~\ref{sec:discussion} below). Since the DPR scales linearly with this parameter, we note that the DPR quoted in Table \ref{tab:progenitor_properties} must be multiplied by five to compare it with the DPR value from \citet{Kilpatricketal2023}, $1.3 \pm 0.1 \times 10^{-8}$ M$_\odot$~yr$^{-1}$. The latter value is therefore $\sim 20\times$ lower than ours, partly because they assumed graphitic dust (see Section \ref{subsec:comparison} for more about this assumption), whose higher opacity results in a lower dust mass, in order to reproduce the observed mid-IR flux.

Since several studies employed the dust radiative-transfer code {\tt DUSTY} \citep{Ivezic1997,Ivezic1999,Elitzur2001} in their analyses of the progenitor candidate \citep{Kilpatricketal2023,Qin2023,Neustadt2024,Xiang2024}, we have also modeled the star in a similar fashion. The primary goal of doing so is not for direct comparison with these other investigations {\it per se}, but as a confirmation of our results with the {\tt GRAMS} modeling. Here we have assumed for the central source the {\tt PHOENIX} model photospheres employed for the {\tt GRAMS} models (rather than the \citealt{Gustafsson2008} {\tt MARCS} atmospheres used by the other studies). Also, similar to our {\tt GRAMS} modeling, we have assumed the \citet{Ossenkopfetal1992} ``warm'' silicates for the dust and the same spherically symmetric dust-shell configuration; additionally, we assumed the ``modified MRN'' \citep{Mathis1977} dust-grain distribution. The grid of models we considered was quite coarse, since our intent here was only to present an approximate comparison to our {\tt GRAMS} results and not necessarily provide a robust fit to the data: The model photosphere steps in $T_{\rm eff}$ were 200~K, the steps in $T_{\rm in}$ (the inner temperature of the dust shell) were 100~K, and the optical depth at 0.55~$\mu$m, $\tau_{\rm V}$, was considered in steps of unity. We also employed only a simple chi-squared, $\chi^2$, goodness-of-fit. We similarly excluded the observed $K_s$ datapoint in the fitting.

\begin{figure*}[ht!]
\plottwo{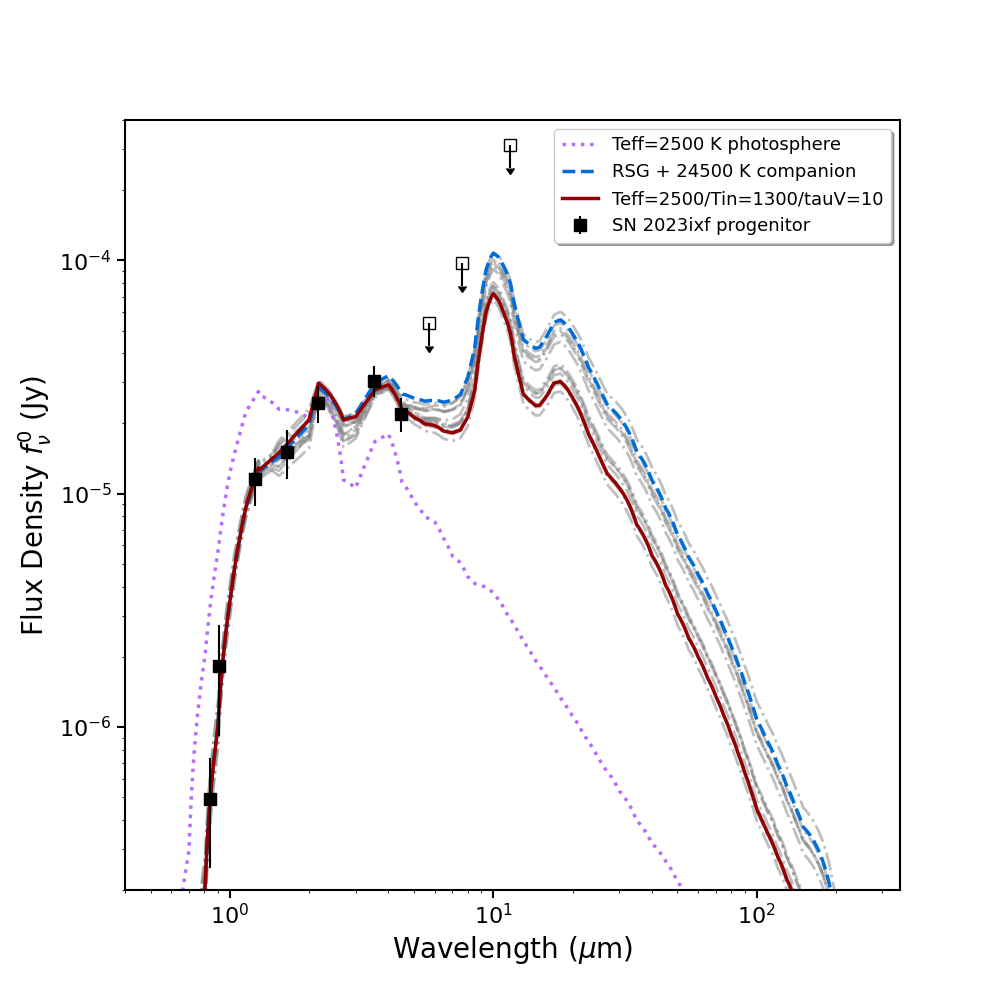}{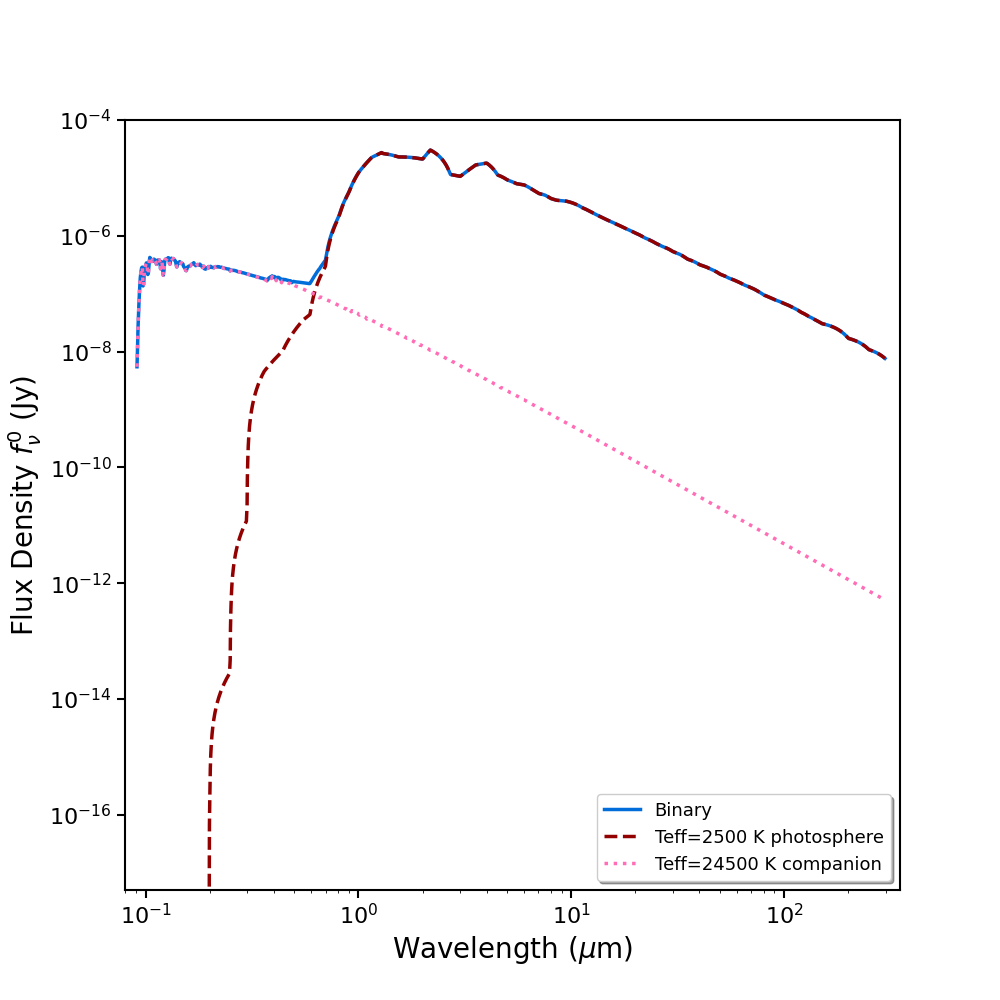}
\caption{Analysis of the SN 2023ixf progenitor candidate SED using {\tt DUSTY} \citep{Ivezic1997,Ivezic1999,Elitzur2001}. {\it Left}: The reddening-corrected observed SED (black squares), together with the best-fitting {\tt DUSTY} model (solid red curve), with input parameters $T_{\rm eff}=2500$~K, $T_{\rm in}=1300$~K, and $\tau_V=10$, and a sample of other good-fitting models (gray curves; see text). The $T_{\rm eff}=2500$~K {\tt PHOENIX} photosphere (dotted purple curve) at the inferred $L_{\rm bol}$ (see text) is also shown for comparison. The error bars include a combination of the measurement uncertainties, as well as the range in flux in each band as a result of source variability. Unlike Figure~\ref{fig:GRAMSfit}, only flux upper limits in the {\sl Spitzer\/} IRAC 5.8 and 8.0 $\mu$m and {\sl WISE} W3 bands are shown, since these are the most constraining. The model predictions are consistent with these limits. The blue curve represents a {\tt DUSTY} model computed with a central source comprised of the $T_{\rm eff}=2500$~K RSG photosphere, plus a hot 24,500~K source (a \citealt{Castelli2003} B1V model atmosphere, meant to approximate a putative binary companion) with $\sim 0.14 \times$ the inferred $L_{\rm bol}$ of the best-fitting RSG model (see right panel), assuming the same input parameters. {\it Right}: The input SED to {\tt DUSTY} for the putative binary system (solid blue curve), consisting of the $T_{\rm eff}=2500$~K RSG photosphere (dotted red curve, in this panel) as the primary and the less luminous, hot (24,500~K) companion (dashed magenta curve).
\label{fig:DUSTYfit}}
\end{figure*}

The results are shown in Figure~\ref{fig:DUSTYfit}. The best-fitting model is obtained from inputs $T_{\rm eff}=2500$~K, $T_{\rm in}=1300$~K, and $\tau_{\rm V}=10$, although a number of other models also providing reasonable fits have input ranges of $T_{\rm eff}=2300$--2700~K, $T_{\rm in}=900$--1300~K, and $\tau_{\rm V}=9$--11. Models outside of these ranges provided significantly poorer fits. The luminosity of the best-fitting model is $L_{\rm bol}=8.2 \times 10^4\ L_{\odot}$, which is somewhat less than the MLE and median values, however, certainly within the credible interval, for the {\tt GRAMS} models. The $T_{\rm eff}$ is somewhat lower than the {\tt GRAMS} MLE and median, although certainly within the credible interval. The $\tau_{\rm V}=10$ corresponds to ${\tau}_{1}=2.9$ and ${\tau}_{10}=0.9$, which are both somewhat larger than the {\tt GRAMS} values, although the latter optical depth is just within the credible interval. The inferred $R_{\rm in}/R_{\star}$, resulting from $T_{\rm in}$, $L_{\rm bol}$, and $T_{\rm eff}$, is 3.4, with $R_{\star}\approx 1534\ R_{\odot}$. Whereas $R_{\rm in}/R_{\star}$ is significantly smaller than the credible interval from the {\tt GRAMS} modeling, $R_{\star}$ itself agrees with the {\tt GRAMS} results. Despite the inputs and assumptions for both the {\tt GRAMS} and {\tt DUSTY} modeling being essentially the same, we found some differences as described here. \citet{Ueta2003} compared {\tt {\bf 2D}UST}, the source routine behind the {\tt GRAMS} models, with {\tt DUSTY} and found overall good agreement in the results, although the differences in the treatment of geometry, particularly at the inner shell edge, likely accounted for some minor discrepancies. Overall, though, whether we had singularly relied on {\tt GRAMS} or {\tt DUSTY}, we ultimately would have arrived at quite similar $T_{\rm eff}$ and $L_{\rm bol}$ estimates for the star.

\begin{figure}[ht!]
\plotone{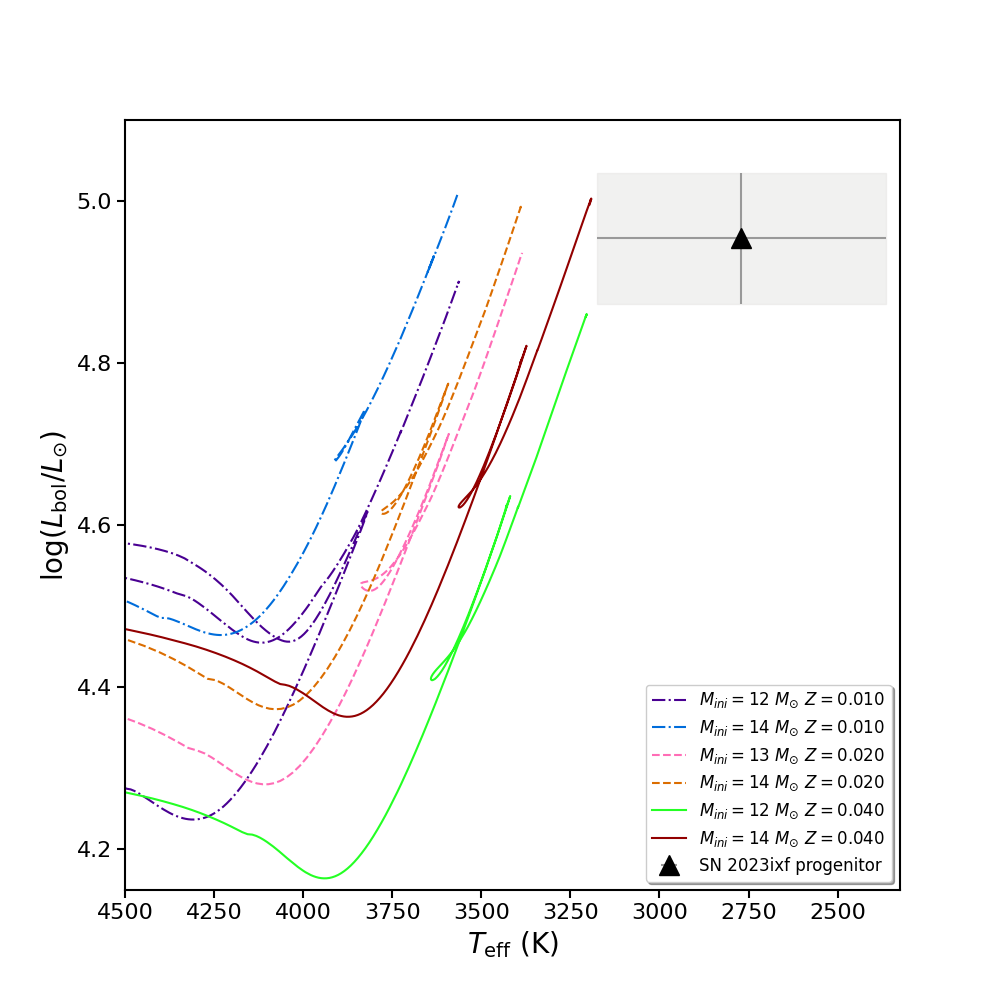}
\caption{Hertzsprung-Russell (HR) diagram showing the locus of the progenitor candidate  based on the median (solid triangle), together with the 68\% credible interval of values (shaded region), from Table~\ref{tab:progenitor_properties}. For comparison we also display single-star theoretical evolutionary tracks from \citet{Stanway2018} at solar ($Z=0.020$; dashed line), subsolar ($Z=0.010$; dash-dot line), and supersolar ($Z=0.040$; solid line) metallicities.
\label{fig:HRD}}
\end{figure}

In Figure~\ref{fig:HRD} we present a Hertzsprung-Russell (HR) diagram with the locus of the progenitor candidate defined by the 68\% credible interval of values for $T_{\rm eff}$ and $L_{\rm bol}$ from Table~\ref{tab:progenitor_properties}. For comparison we show single-star theoretical evolutionary tracks from \citet{Stanway2018} at solar ($Z=0.020$), subsolar ($Z=0.010$), and supersolar ($Z=0.040$) metallicities. One can see that the higher-metallicity supersolar tracks provide better agreement with the progenitor candidate's location in the HR diagram, given that these models have a correspondingly cooler Hayashi limit. Based on the endpoints of these models (we note that the termini for these BPASS tracks is at the end of carbon burning), we can infer that $M_{\rm ini}$ for the progenitor candidate ranges from $12\ M_{\odot}$ to as high as $14\ M_{\odot}$, depending on metallicity. These model stars all terminate within a range of effective radii $R_{\rm eff} \approx 740$--$1028\ R_{\odot}$.

We also considered BPASS binary models, at solar, subsolar, and supersolar metallicity, for which the primary in the system terminates with the appropriate range in $L_{\rm bol}$ and with $T_{\rm eff} \leq 3700$~K. Furthermore, we imposed the criteria from \citet{Eldridge2013,Eldridge2017} for the primary to end as an SN~II-P: total primary mass $M_{\rm prim} \geq 1.5\ M_{\odot}$, CO core mass $\geq 1.38\ M_{\odot}$, ONe core mass $\geq 0.1\ M_{\odot}$, total H mass $> 1\ M_{\odot}$, and the ratio of total H-to-He mass $> 1.05\ M_{\odot}$. We eliminated from consideration model systems, all with initial periods $<4$ d, for which the orbital separation wound up well inside the primary's envelope at terminus, with the remainder being wide binaries with initial masses for the primary of $M_{\rm prim}=12$--$14\ M_{\odot}$ (up to $15\ M_{\odot}$ for the supersolar models), initial periods between $\sim 1000$ and $\sim 10,000$~d (orbital separations from $\sim 1150$ to $\sim 5968\ R_{\odot}$) over a wide range of initial companion-to-primary mass ratios $q=M_{\rm comp}/M_{\rm prim}$. The evolution of the primary was essentially unaffected by the presence of the companion, and the primary's track was the same as if it were a single star. The final separations are generally in the range of 1.1--$7.7\ R_{\star}$. However, for just a few allowed model tracks, the primary is in a long-period system, and the star appears to lose more mass during its lifetime than it would if it were single (based on the BPASS $\dot M$ prescription), while the companion gains some mass, presumably through some level of interaction with the companion. Although the details of these models may not fully apply, they are at least suggestive that a (close) binary companion could have induced additional mass loss from the primary at some point prior to explosion, which could be relevant for the SN 2023ixf progenitor.

\section{Conclusions and Discussion}\label{sec:discussion}

\subsection{Summary}

In this paper we have analyzed the combination of the ground-based near-IR and {\sl Spitzer\/} data from Paper I, together with the available {\sl HST\/} data, on the SN~2023ixf progenitor candidate in M101. The only other hybrid, space- and ground-based SN~II-P progenitor identifications of which we are aware  have been for the nearby Type II SN~2003gd \citep{VanDyk2003,Smartt2004} and SN~2004et \citep{Li2005,Crockett2011}. Quite notably, unlike what has been the case in the past for a large number of progenitor identifications, the {\sl HST\/} pre-SN data were of comparatively little value in understanding the progenitor candidate's nature, since the overwhelming majority of the star's emission prior to explosion was emerging in the IR. We confirmed that the progenitor candidate at {\sl HST\/} F814W is most likely associated with the SN, via observations of SN 2023ixf with the `Alopeke instrument at Gemini-North, and also found that the {\sl HST\/} detection is most likely the counterpart of the detection at 2 $\mu$m (additionally, the near-IR detections are the counterparts of the {\sl Spitzer\/} detections).

We have confirmed that the reddening internal to the host galaxy is likely quite low, via measurements of features, in particular Na~{\sc i}~D, in a high-resolution Keck HIRES spectrum of the SN; this conclusion is also supported by the early-time $B-V$ color of the SN, after correction for the Galactic foreground. We have adopted a total visual extinction $A_V=0.12$ mag, including the Galactic foreground. We have also assessed a likely value for the metallicity at the SN site, based on the spectrum of nebular emission at the outskirts of the catalogued H~{\sc ii} region H1086 $\sim 8{\farcs}3$ northwest of the site, and concluded from various strong-line indicators that it was likely somewhat subsolar ($Z=0.010$) to solar ($Z=0.020$).

We have employed dust radiative-transfer {\tt GRAMS} models to fit the observed SED of the progenitor candidate, corrected first for the total reddening. We find that the star is heavily dust-obscured (high $\tau$) from a likely dusty circumstellar shell or shells. The properties of the star correspond to a median of $T_{\rm eff}=2770$~K and $L_{\rm bol}=9.0 \times 10^4\ L_{\odot}$, with 68\% credible intervals of 2340--3150~K and (7.5--10.9) $\times 10^4 \ L_{\odot}$, respectively. The 68\% credible interval for the DPR is (4.0--12.8) $\times 10^{-8} \ M_{\odot}$ yr$^{-1}$. We have also performed {\tt DUSTY} modeling of the observed SED and found consistent values for $L_{\rm bol}$ and $T_{\rm eff}$; whether we approached the modeling via {\tt GRAMS} or {\tt DUSTY}, we arrived at the same overall picture of the progenitor candidate's nature.

As we had stated in Section~\ref{sec:properties}, the low values of $T_{\rm eff}$ resulting from the modeling are not well-constrained, as a result of the large dust obscuration of the star. An illustrative example applies to the extreme dusty RSG, VY Canis Majoris, for which $T_{\rm eff}$ from modeling results in $\sim 2800$~K \citep[e.g.,][]{Monnier1999}, whereas the luminosity and diameter estimates from interferometry imply $\sim 3500$~K \citep[e.g.,][]{Wittkowski2012}.

We have placed the median values, together with the 68\% credible intervals, for $T_{\rm eff}$ and $L_{\rm bol}$ on a Hertzsprung-Russell diagram and have compared these to the endpoints of BPASS single-star models at both solar and subsolar metallicities, concluding that the progenitor candidate likely had $M_{\rm ini} = 12$--$14\ M_{\odot}$, depending on metallicity. This mass range is consistent with the results of hydrodynamical modeling of the bolometric light curve by \citet{Bersten2024}. Binary models are also possible; however, the overwhelming majority of these models correspond to long-period, wide binaries, in which the primary evolves essentially as a single star, with little effect on it from its binary companion.

\subsection{Comparison With Previous Studies}\label{subsec:comparison}

Our inferred range in this paper for $M_{\rm ini}$, based on the SED modeling, is significantly lower than what we estimated in Paper I, based on the RSG period-luminosity relation: $M_{\rm ini} = 20 \pm 4\ M_{\odot}$. We can also compare with the estimates of $M_{\rm ini}$ for the progenitor candidate in the previous studies by other investigators. \citet{Pledger2023} found a low $M_{\rm ini} \approx 8$--$10\ M_{\odot}$; however, their estimate was entirely based on the available {\sl HST\/} data, whereas we have shown that the vast majority of the flux from the star must be emerging in the IR. We do note that our value at {\sl HST\/} F814W agrees, to within the uncertainties, with that of \citet[][ as well as with \citealt{Soraisam2023a} and \citealt{Kilpatricketal2023}]{Pledger2023}. \citet{Kilpatricketal2023} also concluded that $M_{\rm ini}$ was comparatively low, at $11\ M_{\odot}$. Those authors modeled the candidate assuming {\tt MARCS} atmospheres as input and employing {\tt DUSTY}, which led them to a hotter $T_{\rm eff} \approx 3920$~K and a lower $L_{\rm bol} \approx 10^{4.74}\ L_{\odot}$ than the results of our modeling. 

Although \citet{Kilpatricketal2023} also used {\tt Dolphot}, even after converting their {\sl HST\/} measurements of detections from AB to Vega magnitudes, our measurements were generally systematically brighter (by as much as $\sim 1.7$ mag, in the case of the 1999 F675W observation); interestingly, our F814W measurement agrees with theirs, to within the uncertainties --- see Table~\ref{tab:hst_obs} (cf.~\citeauthor{Kilpatricketal2023}, their table 1). We cannot provide any definitive explanation for these differences, since there are a number of input factors that may have led to them. We also cannot explain the differences in the upper limits to detections, although these differences are not extreme --- as we explained above, we have based our limits on 5$\sigma$ {\tt Dolphot} detections in the overall vicinity of the SN site, whereas their threshold was not explicitly spelled out in their paper. Note that we detected a source at the SN position in the 2014 F673N observation, whereas they did not; and, we considered the 2014 F502N data, and they had not. In Paper I we already described the differences of our measurements in the IR with those by \citet{Kilpatricketal2023}, which may also work to contribute to the differences in $L_{\rm bol}$ and $M_{\rm ini}$.

We also found differences between our {\tt Dolphot} photometry and that by \citet{Niuetal2023}. The values in that study all tended to be fainter than ours --- at the extreme, their detection at F675W differed by $\sim 0.9$ mag --- and this also applied to the upper limits (although those differences, again, were not severe). \citeauthor{Niuetal2023}~also considered far less of the available archival {\sl HST\/} data than we (or \citealt{Kilpatricketal2023}) did. We stand generally by our {\tt Dolphot} results, since our methods and procedures have been developed over the years in consultation with the {\tt Dolphot} author and following the techniques of large {\sl HST\/} photometric surveys (see \citealt{VanDyk2017}).

The {\sl Spitzer\/} flux densities measured by \citet{Niuetal2023} are on average $\sim 13$ and $\sim 12$ $\mu$Jy at 3.6 and 4.5 $\mu$m, respectively (or $\sim 0.5$ mag in both bands), fainter than the values we measured. We note that \citeauthor{Niuetal2023} performed PSF-fitting photometry with {\tt DoPHOT} on the post-processed (mosaicked) Basic Calibrated Data; no details were provided regarding calibration of the photometry or how the PSF was modeled. In general, attempting point source fitting photometry of the IRAC mosaics is not recommended, since the mosaicking process both blurs the undersampled point sources and loses the pixel-phase information\footnote{See https://irsa.ipac.caltech.edu/data/SPITZER/\newline{docs/irac/iracinstrumenthandbook/58/\#{\textunderscore}Toc82083754.}}. PRF fitting with {\tt APEX} of the individual (corrected) Basic Calibrated Data frames, as we have implemented both in Paper I and in this paper, can, with the proper corrections applied, provide photometric measurements to $<1$\%\footnote{See https://irsa.ipac.caltech.edu/data/SPITZER/\newline{docs/irac/iracinstrumenthandbook/57/\#{\textunderscore}Toc82083747.}}.

Both \citet{Kilpatricketal2023} and \citet{Niuetal2023} use pure-carbon best-fit models. A justification for carbonaceous dust grains in RSG circumstellar shells arises from the detection of polycyclic aromatic hydrocarbons (PAHs) in their mid-IR spectra and the presence of continuum emission in the 3--8 $\mu$m range, in excess of that predicted by silicate models \citep{Verhoelst2009}. The availability of C is due to the dissociation of CO by chromospheric UV photons \citep{Becketal1992}. \citet{Verhoelst2009} employed a dust composition with a small mass fraction ($\lesssim 5$\%) of amorphous C to fit the SEDs in their sample, which would require $\lesssim 0.1$\% of the CO to be dissociated. Pure-C dust models require a much larger dissociation fraction ($\sim 10$\%), which is not supported by models of chromospheric dissociation \citep[see, e.g.,][ their figure 1]{Becketal1992}. Moreover, this estimate requires the unrealistic assumption that all of the dissociated C is locked into dust. \citet{Verhoelst2009} admitted that an alternate solution to the problem is to assume that large ($>0.1\ \mu$m) silicate grains can form in these atmospheres, which would reproduce the flux at wavelengths shorter than 8 $\mu$m. Subsequent work \citep[e.g.,][]{Hoefner2008} has shown that this is indeed possible, and such grains have been detected in Galactic RSGs \citep{Sciclunaetal2015, Sciclunaetal2020}. In light of this simpler solution to the problem, the justification for pure-C models is unfounded.

Our results are similar to those of \citet{Jencson2023}, who also used the {\tt GRAMS} models (although, with a different fitting method), as well as a ``super-wind'' model, and found $T_{\rm eff} \approx 3500$~K and $L_{\rm bol} \approx 10^{5.1}\ L_{\odot}$, which, together with the inferred $M_{\rm ini} =17 \pm 4\ M_{\odot}$, are essentially consistent with our values, to within the uncertainties. \citet{Jencson2023} only included the {\sl HST\/} photometry from \citet{Pledger2023} in their SED fitting. We described in Paper I that our {\sl Spitzer\/} measurements at 4.5 $\mu$m agreed well with those by \citet{Jencson2023}, although disagreement exists at 3.6 $\mu$m, which might have contributed to differences in the fitting and the final progenitor candidate properties. Note that \citet{Choi2016} pointed out that, at nominally solar metallicity, the predicted slopes of the MESA Isochrones and Stellar Tracks (MIST) RSG tracks are too shallow compared to the observations; hence, unlike \citeauthor{Kilpatricketal2023} and \citeauthor{Jencson2023}, we did not use those tracks here.

The \citet{Niuetal2023} models are constructed using assumptions and parameter values very similar to ours (thick dust shells, comparable $L_{\rm bol}$, $T_{\rm eff}$, $R_{\rm in}$, and $\tau$). The main discrepancies between our results arise from four sources. (1) \citeauthor{Niuetal2023}~included the F675W detection (which, we have already pointed out, is fainter than ours) in their SED fitting, which we did not. (2) Those authors rejected two of their silicate models based on the temperatures not being consistent with the stellar track endpoints; even for Galactic RSGs with \emph{spectro-interferometric\/} constraints on the radius and temperature of the sources, RSGs are often found at temperatures beyond the termini of single-star evolutionary tracks, and so this should not be considered an effective way to exclude parameter space \citep[see, e.g.,][ their figure 10]{Wittkowskietal2017}. Indeed, the 68\% credible intervals we find for the $T_{\rm eff}$ and $L_{\rm bol}$ of the progenitor lie close to those of two stars, V602~Car and HD~95687, in \citet{Wittkowskietal2017}. (3) \citeauthor{Niuetal2023}~accounted for the variability by adding a 0.5 mag uncertainty to the optical fluxes, which is much lower than the value we use, therefore assigning a higher relative weight to the optical data points when computing their best fit. This inherently restricts the range of IR fluxes that can be probed by the best-fit model. (4) \citeauthor{Niuetal2023}~rejected the remaining silicate models in favor of the pure-C models, because the predicted 8 $\mu$m flux is not lower than their estimated upper limit (which, incidentally, is lower than ours, as a result of the different photometric techniques and assumptions), and those authors estimated a low probability of this being the case. As discussed above, we believe that the justification for a pure-C model is far weaker than for silicate models. Had those authors chosen the silicate models, their results, in terms of the range of possible $M_{\rm ini}$ might well have been more consistent with our inferred range.

Rather unintentionally, our estimates for $M_{\rm ini}$ wound up being roughly consistent with the $\sim 15\ M_{\odot}$ which \citet{Szalai2023} hastily rendered shortly after the SN discovery.

\subsection{Further Thoughts}

\begin{figure}[ht!]
\plotone{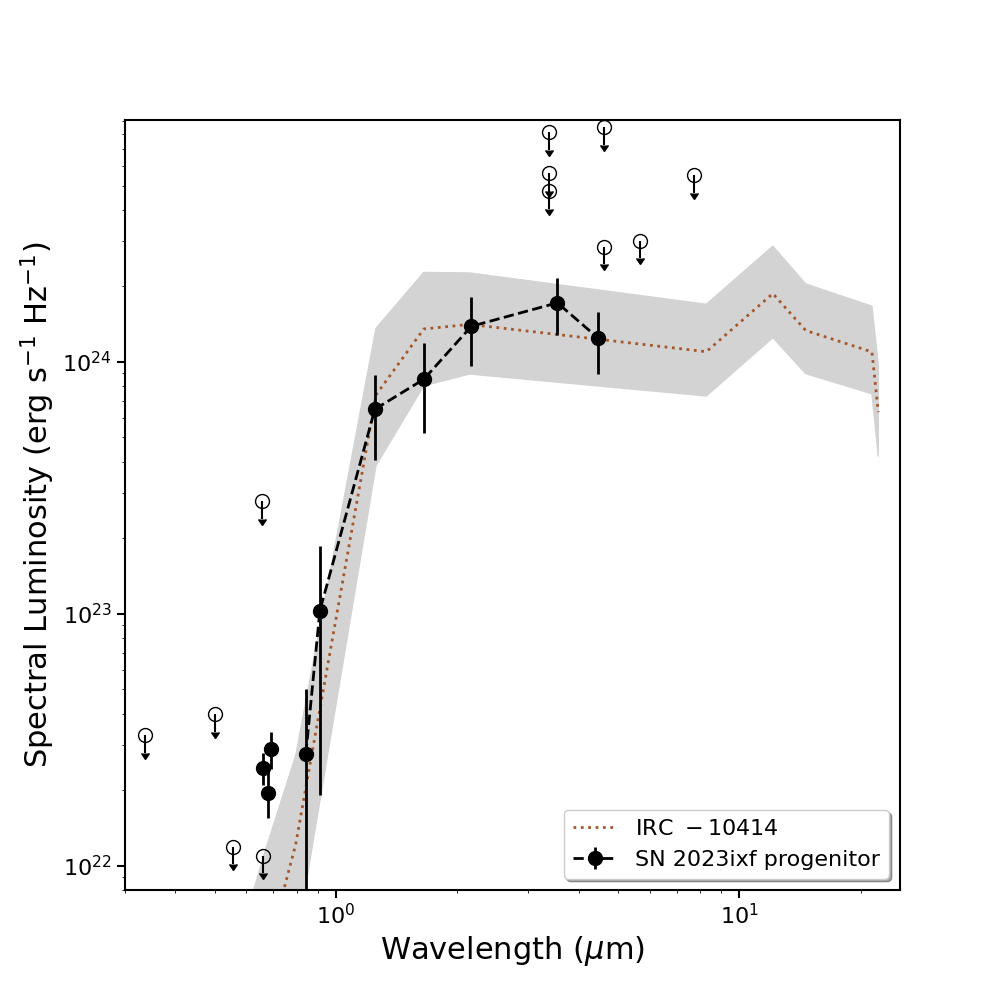}
\caption{A direct comparison of the absolute SED for the SN 2023ixf progenitor candidate (black points; at the assumed distance, see Section~\ref{sec:distance}) with that of the Galactic RSG IRC~$-$10414 (dashed brown curve, with 1$\sigma$ uncertainties as the gray region; \citealt{Gvaramadze2014,Messineo2019,Healy2023}; S.~Van Dyk, in prep.). The implications of the analyses represented in both panels is that the progenitor candidate had luminosity $L_{\rm bol} \approx 10^{5.0}\ L_{\odot}$ in the years prior to explosion.
\label{fig:GalacticRSGs}}
\end{figure}

As something of a confirmation of our modeling results, we show in Figure~\ref{fig:GalacticRSGs} a direct comparison of the progenitor candidate SED with that of the Galactic RSG IRC~$-$10414, a luminous late-M star and an OH, H$_2$O, and SiO maser source (see \citealt{Gvaramadze2014}, and references therein). Additionally, it is also known to be variable with a period $\sim 768$~d\footnote{See https://www.astrouw.edu.pl/asas/ and enter ``182318-1342.8'' under ``ACVS/variables''.}. The IRC~$-$10414 SED is compiled from \citet{Messineo2019} and part of a larger set (S.~Van Dyk, in preparation). The two SEDs are strikingly similar, to within the uncertainties. \citet{Gvaramadze2014} inferred $T_{\rm eff}=3300$~K and \citet{Messineo2019} found a similar, but somewhat cooler, $3110 \pm 170$~K for the Galactic star. The bolometric magnitude we adopt here is $M_{\rm bol}\approx -7.75$, and assuming $M_{\rm bol}(\odot)=4.74$ mag \citep{Mamajek2015}, this corresponds to $L_{\rm bol}\approx 10^{5.00}\ L_{\odot}$, consistent with the estimated luminosity from the {\tt GRAMS} model fitting. (Note that \citealt{Gvaramadze2014} estimated a higher luminosity for the star, $L_{\rm bol}\approx 10^{5.2}\ L_{\odot}$, based on different assumptions.) Interestingly, like the luminous Galactic RSGs Betelgeuse ($\alpha$~Ori; \citealt{NoriegaCrespo1997}) and $\mu$~Cep \citep{Cox2012}, IRC~$-$10414 is associated with an interstellar bow shock \citep{Gvaramadze2014}.

We can make a comparison of the SN 2023ixf progenitor candidate with the two confirmed dusty RSG progenitors of SN 2012aw \citep{VanDyk2012b,Fraser2012} and SN 2017eaw \citep{Kilpatrick2018,VanDyk2019,Rui2019}, as well as of SN 2018aoq \citep{ONeill2019}, which has been shown not to require circumstellar dust in the progenitor SED fitting. For SN 2012aw, \citet{Kochanek2012} presented a different treatment of the dust and constrained the progenitor luminosity to $10^{4.8} < (L_{\rm bol}/L_{\odot}) < 10^{5.0}$. \citet{VanDyk2019} estimated the luminosity of the SN 2017eaw progenitor as $1.2 (\pm 0.2) \times 10^5\ L_{\odot}$, and $M_{\rm ini} \approx 15\ M_{\odot}$. Here we have adjusted the distance to the SN 2017eaw host, NGC~6946, from $7.73 \pm 0.78$ Mpc \citep{VanDyk2019} to the more recent and likely superior estimate, $7.12 \pm 0.38$ Mpc \citep{Anand2021}. The adjusted luminosity would then be $1.02 \times 10^5\ L_{\odot}$ (which also then reduces the estimated $M_{\rm ini}$ to 13--$14\ M_{\odot}$). Here we have also adjusted the distance to the SN 2018aoq host, NGC~4151, from 18.2 Mpc in \citet{ONeill2019} to, again, the more recent Cepheid-based $15.8 \pm 0.4$ Mpc by \citet{Yuan2020}. The SN 2018aoq luminosity in \citet{ONeill2019} is $L_{\rm bol} \approx 10^{4.7}\ L_{\odot}$. Adjusted, this is $L_{\rm bol} \approx 10^{4.58}\ L_{\odot}$. 

\begin{figure}[ht!]
\plotone{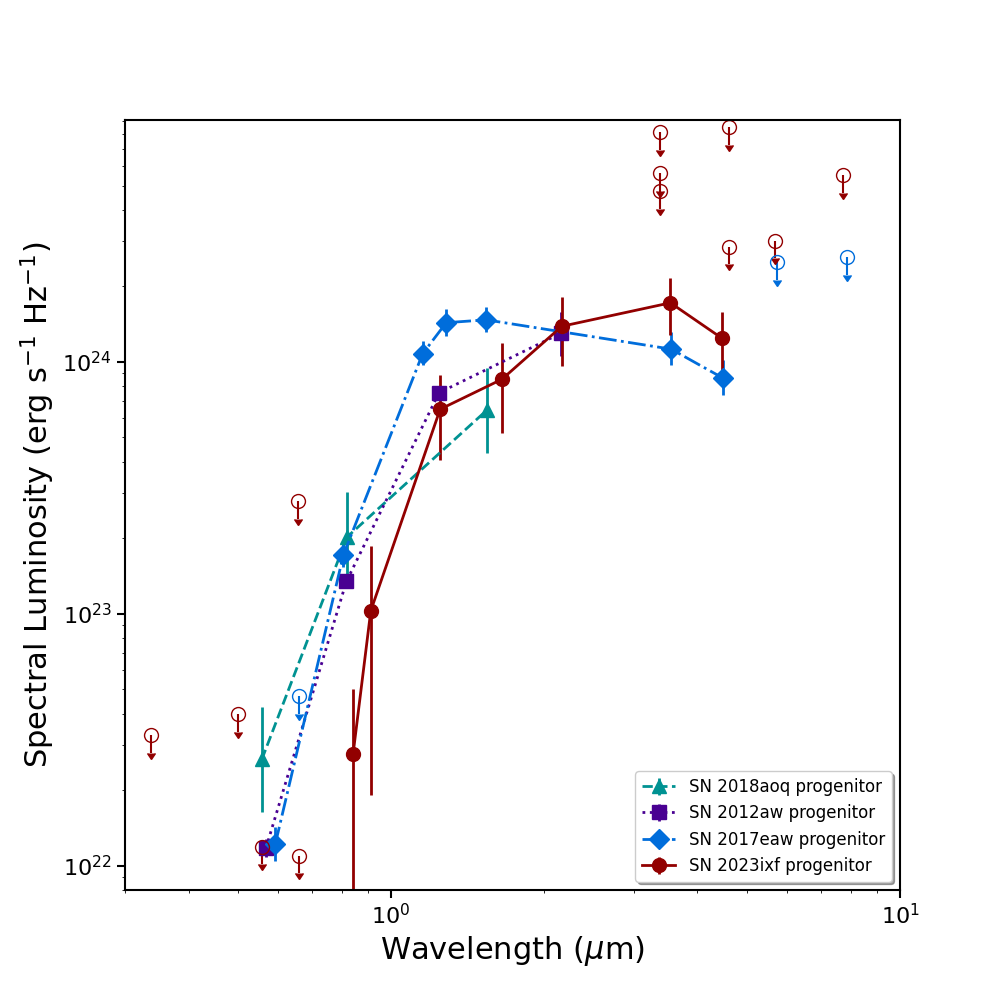}
\caption{A comparison of SN II-P progenitors. We show the reddening- and distance-corrected SED for the SN 2023ixf progenitor (circles), together with the SED for the progenitors of SN 2012aw (squares; \citealt{VanDyk2012b,Fraser2012,Kochanek2012}, SN 2017eaw (diamonds, \citealt{VanDyk2019}; see also \citealt{Kilpatrick2018,Rui2019}), and SN 2018aoq (triangles, \citealt{ONeill2019}). For the latter two SN progenitors, we have adjusted the published luminosities by more recent measurements of the distances to the host galaxies (see text).
\label{fig:sne_comp}}
\end{figure}

We show the overall comparison in Figure~\ref{fig:sne_comp}. It is evident that the SN 2023ixf progenitor candidate would be the dustiest progenitor we as a community have encountered so far. The optical emission from the star is highly suppressed, relative to the other progenitors in the comparison. Correspondingly, as the UV/optical light was reprocessed, far more emission emerged in the IR out to 4.5 $\mu$m, and we would further expect a much higher luminosity than the other stars at wavelengths redward of that (as the SED modeling implies). The total bolometric luminosities of the SN 2023ixf progenitor candidate and the SN 2012aw and 2017eaw progenitors are quite similar, although the difference in the shapes of the three SEDs appears to be stark. How can these otherwise similarly luminous SN progenitors be so different? How this larger total dust opacity, given the likelihood of similar dust stoichiometry \citep[e.g.,][]{Verhoelst2009}, arose for the SN 2023ixf candidate is unknown, but we can speculate that these differences correspond to a more massive and potentially larger circumstellar environment set up around SN 2023ixf, possibly as the result of a higher $\dot M$, stronger wind, due to whatever cause or causes, or binary interaction. 

Our comparison accentuates the heterogeneity in the characteristics of SN progenitors, even those at similar luminosities. This logically parallels that, although general trends exist in observed properties, variations exist between known RSGs, for example in Local Group galaxies. Differences in individual stellar evolution, up to the end of the star's life, are at play, and this could be further affected by the presence and proximity of a binary companion. This comparison emphasizes that each identified SN~II progenitor is valuable and should be individually considered going forward, and that broad-brush conclusions, based on the ensemble, should be made taking all of these considerations into account.

That the progenitor star set up a dense, confined CSM seems extraordinarily likely. How the CSM arose remains uncertain currently. \citet{Hiramatsu2023}, for instance, presented modeling of the pseudobolometric light curve over the first month and concluded that enhanced mass loss from the progenitor must have occurred during the final 1--2~yr before explosion, either through a single eruption or a continuous mass-loss process at $\dot M \approx 0.01$--1.0 $M_{\odot}$ yr$^{-1}$, leading to the formation of CSM of 0.3--1.0~$M_{\odot}$. However, \citet{Neustadt2024} analyzed a roughly 15-yr-long optical dataset (spanning 5600 to 400~d before explosion) obtained with the Large Binocular Telescope (LBT) and found no evidence for pre-explosion outbursts during this period (however, because of the sparse coverage of their data, they could not directly exclude short-lived outbursts). \citet{Ransome2023} also found no significant detections in long-baseline, multiband Pan-STARRS light curves prior to explosion. \citet{Flinner2023} also ruled out luminous UV eruptions 15--20 yr prior to explosion. \citet{Dong2023} examined over $\sim 5$~yr of pre-explosion data from the DLT40, ZTF, and ATLAS surveys and also concluded that the probability of precursor outbursts is low. Again, being insensitive to short outbursts, those authors set upper limits of 100 and 200~days for the duration of any such outbursts in the case of a peak brightness of $M_r \approx -9$ mag and $-8$ mag, respectively, leading to a maximal amount of ejected pre-explosion matter of $0.015\ M_{\odot}$. The formation mechanism, or mechanisms, for the CSM remains elusive. We note, however, that we have shown that it could be plausible to set up the CSM with just a steady, low wind velocity and a moderate $\dot M$ over the course of the RSG phase, without the need to invoke a super-wind or eruption: the range in values for $R_{\rm in}$ for the dust shell are essentially consistent with the estimates of the confined CSM dimensions obtained from the SN itself \citep[e.g.,][]{Jacobson-Galan2023,Bostroem2023}. Furthermore, \citealt{Zimmerman2024} argued that radiative acceleration can explain the observed high-velocity flash-feature profiles without invoking a recent pre-SN stellar eruption to accelerate matter. We caution that wind velocities $\gtrsim 100$ km s$^{-1}$ might even hamper significant dust production, since matter needs to remain long enough within the dust-formation zone.

It is intriguing to suggest that IRC~$-$10414 may serve as a Galactic analog for the progenitor of SN 2023ixf, not that we are stating here that the progenitor candidate was a maser source --- we obviously have no way of confirming that possibility --- nor that IRC~$-$10414 is only decades, years, or days, from core collapse. However, in the context of this possible analog, particularly notable is the flux excess in the {\sl HST\/} bands shortward of 8000~\AA\ (F658N, F673N, and F675W), relative to the overall fit to the observed SED of the candidate including the flux at F814W, and to the IRC~$-$10414 SED (see Figure~\ref{fig:GalacticRSGs}). The first two narrow bands are sensitive to H$\alpha$($+$[N~{\sc ii}]) and [S~{\sc ii}], respectively, while the broader F675W (WFPC2 $\sim R$) would be sensitive to detecting both. The excess could imply that line emission is present within the PSF of the progenitor candidate, not necessarily from the star itself, but from its immediate environment. It is interesting to note that \citet{Gvaramadze2014} detected strong H$\alpha$$+$[N~{\sc ii}], and comparatively weaker [S~{\sc ii}], from the arc-like bow shock at $\sim 15\arcsec$ ($\sim 0.14$ pc) from IRC~$-$10414; with FWHM $\approx 1.5$ ACS pixel for the candidate profile, at 6.85 Mpc this is $\sim 2.5$~pc, so an analogous bow shock would be within the PSF. If this were a bow shock, it could imply that the SN 2023ixf progenitor candidate possibly had been a runaway star. It is indeed curious that this candidate, as well as the progenitors of SN 2012aw and SN 2017eaw, were all isolated spatially from any obvious stellar clustering, which is not what we would necessarily expect for such an initially massive star.

Other explanations for a possible flux excess in those {\sl HST\/} filters include an ionized circumstellar shell around the progenitor candidate, analogous to what \citet{Wright2014} found around the RSG W26 in the massive Galactic cluster Westerlund~1, or a photoionized confined circumstellar shell, such as that around Betelgeuse \citep{Mackey2014}. Since RSGs are clearly too cool to photoionize matter around them, the possible source of UV photons, in the case of the candidate, could either be a neighboring hot, blue star or from a hot binary companion. As can be seen in Figure~\ref{fig:progID}, a blue star is $\sim 0{\farcs}24$ to the southwest of the progenitor; however, this is $\sim 7.9$ pc away, which may well be too distant. Alternatively, possibly some shock excitation process was at work in the star's circumstellar environment.

As we pointed out in Section~\ref{sec:properties}, binary progenitor systems are theoretically possible and, in the case of the wide binaries, the larger mass ratios ($\geq 0.4$) for the models would correspond to secondaries (companions) which would be hot enough to ionize any circumbinary environment. The main obstacle here is that the BPASS binary orbital separations span $\sim 1.1$--$7.7\ R_{\star}$, whereas for our dust modeling the preferred $R_{\rm in}$ are 6.5--$13.3\ R_{\star}$, i.e., the binary system would most likely exist entirely within the dust shell, based on our results. Any ionization of the circumstellar environment would likely be on the inner parts of the shell and would be heavily obscured (based on the resulting optical depth from our best-fit {\tt GRAMS} model, $A_{{\rm H}\alpha} \approx 4$ mag).

Another curious aspect of the excess optical emission is that the progenitor candidate is no longer detectable in the ACS F658N observation from 2018 March, more than 14~yr after and a factor of $\sim 2.4$ deeper than the ACS F658N observation in 2004 February. The candidate would have decreased in brightness by $\gtrsim 0.9$ mag. (We note that the candidate was not detected in the WFPC2 F656N imaging in 1999 March and June; however, those observations were not as deep and as sensitive as the ACS ones; that excess light was detected contemporaneously in F675W indicates that some H$\alpha$ emission existed at the time.) Either whatever source of excitation that was responsible was curtailed or obscuring dust was suddenly present, possibly in some episodic event, (this would definitely require the H$\alpha$ emission to have been local to the circumstellar environment). Whatever obscuration there might have been had little effect on the mid-IR emission from the candidate, as detected with {\sl Spitzer\/} up to 2018, as well as the near-IR detected from 2007 through 2013 (see Paper I).

One additional aspect regarding the progenitor as a binary follows on from a point raised by \citet{Kilpatricketal2023}. As we mentioned above, a putative companion would have been completely obscured prior to explosion. We agree with those authors that, among the deep nondetections at F336W, F435W, and F555W, the latter is the most constraining --- with additional CSM obscuration of $A_{\rm F555W} \approx 6$ mag based on the {\tt GRAMS} model, the limit would be $M_{\rm F555W} \gtrsim -8.8$ mag. However, the secondaries of all of the allowed model binary systems are easily less luminous than this limit.

An additional constraint on the presence of a binary companion, as suggested by the reviewer of this paper, is to determine at what companion luminosity would the observed SED be detectably modified by the additional UV from the hot star. We could not undertake this test with the {\tt GRAMS} models, since those are all pre-computed and packaged. We could, however, use {\tt DUSTY}, by including progressively hotter and more luminous blue stars to the central RSG source, until the model SED diverged significantly from the observed one. The assumption is that the hot companion is within $R_{\rm in}$, as both the SED and binary models imply, and can indeed effectively be treated as an additional central flux source within {\tt DUSTY} (this, admittedly, may be an oversimplification). We adopted our best-fitting model parameters, as shown in Figure~\ref{fig:DUSTYfit} (left panel), and the allowed BPASS models. The companions were approximated by \citet{Castelli2003} main-sequence model stellar atmospheres; see Figure~\ref{fig:DUSTYfit} (right panel). We found that a companion would have to be quite massive ($q \gtrsim 0.8$), hot ($T_{\rm eff} \gtrsim 24,500$~K), and luminous ($\gtrsim 10^{4.1}\ L_{\odot}$, $\gtrsim 0.139\ L_{\rm prim}$) for it to have had any appreciable effect on the observed SED. As can be seen in Figure~\ref{fig:DUSTYfit}, the increased UV contribution from the companion noticeably increases the IR emission at $\gtrsim 3\ \mu$m; at this companion luminosity the brightness at {\sl Spitzer\/} IRAC $4.5\ \mu$m would be significantly higher than the actual observed ensemble brightness, including variability, in this band. The total luminosity from the dusty system would also be $\gtrsim 14$~\% higher than what we have inferred from the observed progenitor candidate. Therefore, if the RSG progenitor had a companion, we can surmise that it would have had to be cooler, less luminous, and less massive than this. From the BPASS models a surviving companion would have luminosities of, for instance, $\gtrsim -5.3$ and $\gtrsim -4.8$ mag in WFC3/UVIS F275W and F336W.

\begin{figure*}[ht!]
\plotone{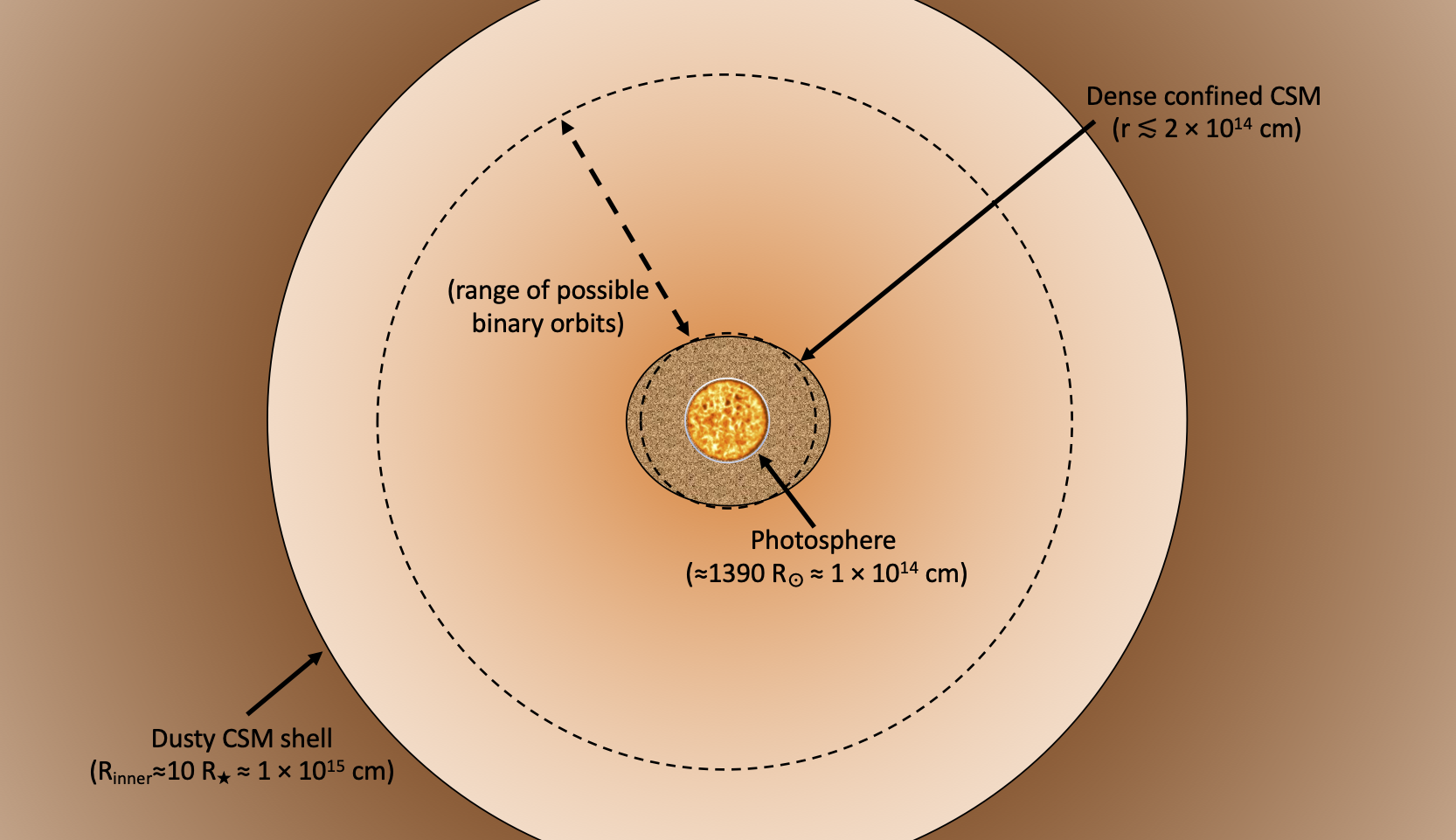}
\caption{A cartoon schematic of the SN 2023ixf progenitor candidate prior to explosion, to approximate scale, based on observational and modeling inference. Here we represent the stellar photosphere at its median effective radius $R_{\star} \approx 1390\ R_{\odot}$. Immediately above that is the dense, asymmetric \citep{Smith2023,Vasylyev2023,Li2023}, confined CSM inferred by \citet{Zimmerman2024}. The inner radius $R_{\rm in}$ of the (assumed) spherically symmetric dust shell, which we infer here from our SED modeling, is shown at its median value of $10\ R_{\star} \approx 1 \times 10^{15}$ cm. The shell is assumed to extend to $1000\ R_{\rm in}$ (not shown), with density decreasing as $\rho \propto r^{-2}$. Additionally, we show the range of allowed model binary orbital radii, which is all within the primary's dust shell (see text).
\label{fig:schematic}}
\end{figure*}

\citet{Zimmerman2024}, via analysis of early-time UV spectra obtained of the SN with {\sl HST\/}, estimated densities and dimensions of different regimes within the star's CSM. Those authors determined that a higher-density ($\sim 5 \times 10^{-13}$ g cm$^{-3}$), confined ($\lesssim 2 \times 10^{14}$ cm; see also \citealt{Martinez2023}) region of CSM above the stellar photosphere (they estimated the star's radius at $\sim 5 \times 10^{13}$ cm, whereas our estimate is a factor of two larger, $\sim 1 \times 10^{14}$ cm; see Section~\ref{sec:properties}) extended the shock breakout. Beyond that the CSM density the density drops (to $\lesssim 10^{-15}$ g cm$^{-3}$) and continues to gradually decline. We have found from our SED modeling that at $\sim 10^{15}$ cm (or, $\sim 10\ R_{\star}$) the CSM was cool enough for dust to form significantly and the resulting dust shell extended outward as the density progressively declined. That \citeauthor{Zimmerman2024}~concluded that the density beyond the dense, confined CSM and shock breakout regime must fall off as $\propto r^{-2}$ lends support to our assumption of a similar density distribution for the dust shell. Following \citet{Grefenstette2023}, \citeauthor{Zimmerman2024}~assumed a gas density of $\sim 4 \times 10^{-16}$ g cm$^{-3}$ at $\sim 10^{15}$ cm, whereas we found the dust density to be $\sim 10^{-17}$ to $\sim 10^{-18}$ g cm$^{-3}$ at this radius ($R_{\rm in}$); these two density estimates are consistent for a reasonable assumed gas-to-dust ratio of 200. We note that the modeling by \citet{Martinez2023} results in a constraint on an extended CSM, at $\sim 8 \times 10^{14}$ cm, which is consistent with the confidence intervals on $R_{\rm in}$; however, we posit that the lower-density dusty CSM extends well beyond that (see also \citealt{Li2023}). In Figure~\ref{fig:schematic} we show a cartoon schematic of the approximate progenitor CSM geometry, with the asymmetric \citep{Smith2023,Vasylyev2023,Li2023}, confined CSM indicated above the photosphere and the inner radius of the simplistically spherical dust shell represented, as well as the range of possible binary orbits; this again illustrates that any binary companion was most likely within the dust shell, and those with the smallest orbits were even potentially within the denser, inner CSM.

The SN should be observed, with either {\sl HST\/} or the {\sl James Webb Space Telescope\/} ({\sl JWST}), when its brightness has decreased enough that the SN's image does not saturate the detectors and when enough fiducial stars around the site also can be imaged at sufficiently high SNR, such that a robust astrometric alignment can be made with the pre-SN imaging data, to more securely associate the SN with the progenitor candidate. \citet{Qin2023} had already attempted this with adaptive optics from the ground, although the SN image was still heavily saturated at the time. Of course, years from now, when the SN has faded to a sufficiently low level (given the indications of CSM interaction, this could be quite a long time), it should be observed again with either of the two space telescopes, to confirm that the progenitor candidate is indeed the actual progenitor. Furthermore, with much of the pre-SN CSM dust destroyed, deep observations in the blue and UV should be performed to detect or place constraints on a binary companion (although, the presence of any remaining pre-existing dust or freshly-formed dust in the SN shock could compromise this set of observations).

With the SN 2023ixf progenitor candidate we have been able to obtain an unprecedented portrait of the potential star that exploded. For one thing, we have a first-ever spectacular view of the semi-regular variability of the progenitor in the years prior to core collapse. For another, we can construct a well-sampled SED from the optical through the mid-IR for the star, which details the final state of the star. Sometime in the not-so-distant future, an SN will occur in a very nearby host galaxy --- possibly even once again M101 --- with unheralded data coverage by both {\sl HST\/} and {\sl JWST\/}, such that a spectacular SED can be constructed for the progenitor star. Until that time, the well-characterized progenitor candidate for SN 2023ixf will have to suffice.

\bigskip
\bigskip
%\begin{acknowledgments}
We are grateful to the reviewer for the several comments that helped improve the manuscript.
We thank Daichi Hiramatsu for providing the photometry for SN 2006Y and SN 2006ai.
This work was authored by employees of Caltech/IPAC under Contract No.~80GSFC21R0032 with the National Aeronautics and Space Administration (NASA).
Based on observations made with the NASA/ESA {\sl Hubble Space Telescope}, obtained from the Data Archive at the Space Telescope Science Institute (STScI), which is operated by the Association of Universities for Research in Astronomy (AURA), Inc., under NASA contract NAS5-26555. 
This work is based in part on archival data obtained with the {\sl Spitzer Space Telescope}, which was operated by the Jet Propulsion Laboratory (JPL), California Institute of Technology under a contract with NASA. Support for this work was provided by an award issued by JPL/Caltech.
This publication makes use of data products from the {\sl Wide-field Infrared Survey Explorer}, which
is a joint project of the University of California, Los Angeles, and JPL/Caltech, funded
by NASA.
We also made use of the NASA/IPAC Extragalactic Database (NED), which is funded by NASA and operated by Caltech.
Based in part on observations obtained at the international Gemini Observatory, acquired through the Gemini Observatory Archive at the National Science Foundation's (NSF's) NOIRLab and processed using {\tt DRAGONS} (Data Reduction for Astronomy from Gemini Observatory North and South). Some of the observations in the paper made use of the High-Resolution Imaging instrument 'Alopeke, which was funded by the NASA Exoplanet Exploration Program and built at the NASA Ames Research Center by Steve B. Howell, Nic Scott, Elliott P. Horch, and Emmett Quigley. 'Alopeke was mounted on the Gemini North telescope of the international Gemini Observatory, a program of NSF's NOIRLab, which is managed by AURA under a cooperative agreement with the NSF, on behalf of the Gemini partnership: the NSF (United States), National Research Council (Canada), Agencia Nacional de Investigaci\'{o}n y Desarrollo (Chile), Ministerio de Ciencia, Tecnolog\'{i}a e Innovaci\'{o}n (Argentina), Minist\'{e}rio da Ci\^{e}ncia, Tecnologia, Inova\c{c}\~{o}es e Comunica\c{c}\~{o}es (Brazil), and Korea Astronomy and Space Science Institute (Republic of Korea). 
Some of the data presented herein were obtained at the W.~M.~Keck Observatory, which is operated as a scientific partnership among Caltech, the University of California, and NASA. The Observatory was made possible by the generous financial support of the W.~M.~Keck Foundation. 
Both Keck and Gemini North are located within the Maunakea Science Reserve and adjacent to the summit of Maunakea. 
The authors wish to recognize and acknowledge the very significant cultural role and reverence that the summit of Maunakea has always had within the indigenous Hawaiian community.  We are most fortunate to have the opportunity to conduct observations from this unique mountain. 
KAIT and its ongoing operation were made possible by donations from Sun Microsystems, Inc., the Hewlett-Packard Company, AutoScope Corporation, the Lick Observatory, the NSF, the University of California, the Sylvia \& Jim Katzman Foundation, and the TABASGO Foundation. Research at Lick Observatory is partially supported by a generous gift from Google.       
This research has made use of the Spanish Virtual Observatory \citep[https://svo.cab.inta-csic.es;][]{SVO2012,SVO2020} project funded by MCIN/AEI/10.13039/501100011033/ through grant PID2020-112949GB-I00.
S.S. acknowledges support from UNAM-PAPIIT Program IA104822.
T.S. is supported by the NKFIH/OTKA grant FK-134432 of the National Research, Development and Innovation (NRDI) Office of Hungary, by the J\'anos Bolyai Research Scholarship of the Hungarian Academy of Sciences, and by the New National Excellence Program (UNKP-22-5) of the Ministry for Culture and Innovation from the source of the NRDI Fund, Hungary.
A.V.F.'s group at U.C. Berkeley is grateful for financial assistance from NASA/HST grant AR-14295, the Christopher R. Redlich Fund, Gary and Cynthia Bengier, Clark and Sharon Winslow, Alan Eustace (W.Z. is a Bengier-Winslow-Eustace Specialist
in Astronomy), and many other donors.
S.H.C. acknowledges support from the National Research Foundation of Korea (NRF) grant funded by the Korea government (MSIT) (NRF-2021R1C1C2003511) and the Korea Astronomy and Space Science Institute under R\&D program (Project No. 2023-1-860-02) supervised by the Ministry of Science and ICT.
%\end{acknowledgments}

\vspace{5mm}
\facilities{HST(WFPC2, ACS, WFC3), Spitzer, WISE, Gemini, Keck:I (HIRES), Lick, Herschel, Akari}

\software{{\tt APEX} \citep{Makovoz2005b}, 
          {\tt AstroDrizzle} \citep{STScI2012}, 
          {\tt astropy} \citep{Astropy_2013,Astropy_2018,Astropy_2022},
          {\tt DAOPHOT} \citep{Stetson1987},
          {\tt Dolphot} \citep{Dolphin2016},
          {\tt DUSTY} \citep{Ivezic1999},
          {\tt DRAGONS} \citep{Labrie2019},
          {\tt emcee} \citep{Foreman-Mackeyetal2013}, 
          {\tt MOPEX} \citep{Makovoz2005a},
          {\tt photutils} \citep{photutils}
          {\tt PyRAF} \citep{Pyraf2012}
          }

%\appendix
%\section{Appendix information}

\bibliography{main}{}
\bibliographystyle{aasjournal}

\end{document}